\documentclass[fleqn,usenatbib]{mnras}

\usepackage{newtxtext,newtxmath}

\usepackage[T1]{fontenc}

\DeclareRobustCommand{\VAN}[3]{#2}
\let\VANthebibliography\thebibliography
\def\thebibliography{\DeclareRobustCommand{\VAN}[3]{##3}\VANthebibliography}

\usepackage{graphicx}	
\usepackage{amsmath}	
\usepackage{subfig}
\usepackage[export]{adjustbox}
\usepackage{comment}
\usepackage{float}

\usepackage{booktabs, multirow} 
\usepackage{soul}
\usepackage{xcolor,colortbl} 
\usepackage{changepage,threeparttable} 
\usepackage{makecell}   

\usepackage{multicol}
\usepackage{enumitem}

\title[UMBH in the Cosmic Horseshoe]{Unveiling a 36 Billion Solar Mass Black Hole at the Centre of the Cosmic Horseshoe Gravitational Lens}

\author[Melo-Carneiro et al.]{Carlos R. Melo-Carneiro,$^{1,2}$\thanks{E-mail: carlos.melo@ufrgs.br}
Thomas E. Collett,$^{2}$
Lindsay J. Oldham,$^{2}$
Wolfgang Enzi,$^{2}$
\newauthor 
Cristina Furlanetto,$^{1}$
Ana L. Chies-Santos,$^{1}$
and Tian Li$^{2}$
\\
$^{1}$Instituto de Física, Universidade Federal do Rio Grande do Sul,
Av. Bento Gonçalves 9500, Porto Alegre-RS, 90040-060, Brazil\\
$^{2}$Institute of Cosmology and Gravitation, University of Portsmouth,
Burnaby Rd, Portsmouth, PO1 3FX, UK\\
}

\date{Accepted XXX. Received YYY; in original form ZZZ}

\pubyear{\the\year{}}

\begin{document}
\label{firstpage}
\pagerange{\pageref{firstpage}--\pageref{lastpage}}
\maketitle

\begin{abstract}
Supermassive black holes (SMBHs) are found at the centre of every massive galaxy, with their masses tightly connected to their host galaxies through a co-evolution over cosmic time. For massive ellipticals, the SMBH mass ($M_\text{BH}$) strongly correlates with the host central stellar velocity dispersion ($\sigma_e$), via the $M_\text{BH}-\sigma_e$ relation. However, SMBH mass measurements have traditionally relied on central stellar dynamics in nearby galaxies ($z < 0.1$), limiting our ability to explore the SMBHs across cosmic time. In this work, we present a self-consistent analysis combining 2D stellar dynamics and lens modelling of the Cosmic Horseshoe gravitational lens system ($z_l = 0.44$), one of the most massive lens galaxies ever observed. Using MUSE integral-field spectroscopy and high-resolution HST imaging, we simultaneously model the radial arc - sensible to the inner mass structure - with host stellar kinematics to constrain the galaxy’s central mass distribution and SMBH mass. Bayesian model comparison yields a $5\sigma$ detection of an ultramassive black hole (UMBH) with $\log_{10}(M_\text{BH}/M_{\odot}) = 10.56^{+0.07}_{-0.08} \pm (0.12)^\text{sys}$, consistent across various systematic tests. Our findings place the Cosmic Horseshoe $\sim$$1.5\sigma$ above the $M_\text{BH}-\sigma_e$ relation, supporting an emerging trend observed in brighter cluster galaxies (BCGs) and other massive galaxies, which suggests a steeper $M_\text{BH}-\sigma_e$ relationship at the highest masses, potentially driven by a different co-evolution of SMBHs and their host galaxies. Future surveys will uncover more radial arcs, enabling the detection of SMBHs over a broader redshift and mass range. These discoveries will further refine our understanding of the $M_\text{BH}-\sigma_e$ relation and its evolution across cosmic time.

\end{abstract}




\section{Introduction}
Most massive galaxies are believed to host a supermassive black hole (SMBH) at their centre. More importantly, host galaxies and their SMBHs exhibit clear scaling relations, pointing to a co-evolution between the galaxy and the SMBH \citep[][]{Kormendy2013}. The SMBH mass ($M_\text{BH}$) has been shown to correlate with various galaxy properties, such as the bulge luminosity \citep[e.g.,][]{Magorrian1998,Marconi2003,Gultekin2009}, stellar bulge mass \citep[e.g.,][]{Laor2001,McLure2002}, dark matter (DM) halo mass \citep[e.g.,][]{Marasco2021,Powell2022}, number of host's globular clusters \citep[e.g.,][]{Burkert2010,Harris2014}, and stellar velocity dispersion \citep[e.g.,][]{Gebhardt2000,Beifiori2009}. Notably, the $M_\text{BH}-\sigma_e$ relation, which links SMBH mass to the effective stellar velocity dispersion of the host ($\sigma_e$), remains tight across various morphological types and SMBH masses \citep{Bosch2016}. Nonetheless, when SMBHs accrete mass from their neighbourhoods, they can act as active galactic nuclei (AGNs), injecting energy in the surrounding gas in a form of feedback. This feedback can be either positive, triggering star formation \citep[][]{Ishibashi2012,Silk2013,Riffel2024}, or negative, quenching galaxy growth \citep[e.g.,][]{Hopkins2006,Dubois2013,Costa-Souza2024}. 

It is expected that the most massive galaxies in the Universe, such as brightest cluster galaxies (BCGs), host the most massive SMBHs. Indeed, so-called ultramassive black holes (UMBHs; $M_\text{BH} \geq 10^{10}M_\odot$) have been found in such systems \citep[e.g.,][]{Hlavacek-Larrondo2012}. Most of these UMBHs have been measured through spatially-resolved dynamical modelling of stars and/or gas. For instance, the UMBH in Holm 15A at $z=0.055$ \citep[$M_\text{BH} = (4.0 \pm 0.80) \times 10^{10}M_\odot$;][]{Mehrgan2019} and the UMBH in NGC 4889 at $z=0.021$ \citep[$M_\text{BH} = (2.1 \pm 1.6) \times 10^{10}M_\odot$;][]{McConnell2012} were both determined using stellar dynamical modelling. However, despite the success of this technique in yielding hundreds of SMBH mass measurements, the requirement for high-quality spatially resolved spectroscopy poses significant challenges for studies at increasing redshift \citep[see, e.g.,][Suplemental Material S1]{Kormendy2013}. 

Nonetheless, the significance of these UMBHs lies in the fact that many of them deviate from the standard linear $M_\text{BH}-\sigma_e$ relation \citep[e.g.,][]{Kormendy2013,Bosch2016}. This suggests either a distinct evolutionary mechanism governing the growth of the largest galaxies and their SMBHs \citep[][]{McConnell2011}, leading to a significantly steeper relation \citep[][]{Bogdan2018}, or a potential decoupling between the SMBH and host galaxy co-evolution. Populating the high-mass end of the $M_\text{BH}-\sigma_e$ relation, particularly through direct $M_\text{BH}$ measurements, could help resolve this ongoing puzzle. 

Recently, \citet{Nightingale2023}, by modelling the gravitationally lensed radial image near the the Abell 1201 BCG ($z=0.169$), was able to measure the mass of its dormant SMBH as $M_\text{BH} = (3.27 \pm 2.12) \times 10^{10}M_\odot$, therefore an UMBH. This provides a complementary approach to other high-$z$ probes of SMBH mass, such as reverberation mapping \citep[][]{Blandford1982,Bentz2015} and AGN spectral fitting \citep[][]{Shen2013}. Unlike these methods, which require active accretion and depend on local Universe calibrations, the lensing technique offers a direct measurement independent of the SMBH's accretion state.

In this paper, we analyse the Cosmic Horseshoe gravitational lens system \citep[][]{Belokurov2007}, where the lens galaxy is one of the most massive strong gravitational lenses known to date. The lens galaxy is an early-type galaxy (ETG) at redshift $z_l=0.44$, possibly part of a fossil group \citep[][]{Ponman1994}, and is notable for lensing one of its sources into a nearly complete Einstein ring (the Horseshoe). Additionally, a second multiply imaged source forms a radial arc near the centre of the lens galaxy. 

Thanks to the radial image formed very close to the centre, the inner DM distribution of the Cosmic Horseshoe can be studied in detail, as done by \cite{Schuldt2019}. By simultaneously modelling stellar kinematics from long-slit spectroscopy and the positions of the lensed sources, \cite{Schuldt2019} found that the DM halo is consistent with a Navarro-Frenk-White \citep[NFW;][]{Navarro1997} profile, with the DM fraction within the effective radius ($R_e$) estimated to be between $60\%$ and $70\%$. Moreover, their models include a point mass at the galaxy's centre, reaching values around $\sim 10^{10} M_\odot$, which could represent an SMBH; however, they did not pursue further investigations into this possibility.

Using new integral-field spectroscopic data from the Multi Unit Spectroscopic Explorer (MUSE) and imaging from the Hubble Space Telescope (HST), we conducted a systematic modelling of the Cosmic Horseshoe system to reassess the evidence for an SMBH at the heart of the lens galaxy. We performed a self-consistent analysis of both strong gravitational lensing (SGL) and stellar dynamics, which demonstrated that the presence of an SMBH is necessary to fit both datasets simultaneously. 

This paper is structured as follows: In Section \ref{sec:data}, we present the HST imaging data and MUSE observations, along with the kinematic maps used for the dynamical modelling. Section \ref{sec:methods} briefly summarises the lensing and dynamical modelling techniques, including the multiple-lens-plane formalism, the approximations adopted in this work, and the mass profile parametrisation. In Section \ref{sec:results}, we present the results from our fiducial model and alternatives models, which we use to address the systematics on the SMBH mass. In Section \ref{sec:discussion} we discuss our results and present other astrophysical implications. Finally, we summarise and conclude in Section \ref{sec:conclusion}.

Unless otherwise, all parameter estimates are derived from the final sampling chain, with reported values representing the median of each parameter's one-dimensional marginalised posterior distribution, with uncertainties corresponding to the $16^\text{th}$ and $84^\text{th}$ percentiles. Furthermore, throughout this paper, we adopt the cosmological parameters consistent with \cite{Planck_2015}: $\Omega_{\Lambda,0} = 0.6911$, $\Omega_{\text{m},0} = 0.3089$, $\Omega_{\text{b},0} = 0.0486$, and $H_0 = 67.74$ km s$^{-1}$ Mpc$^{-1}$.

\section{Data}\label{sec:data}
\begin{figure}
	\includegraphics[width=\columnwidth]{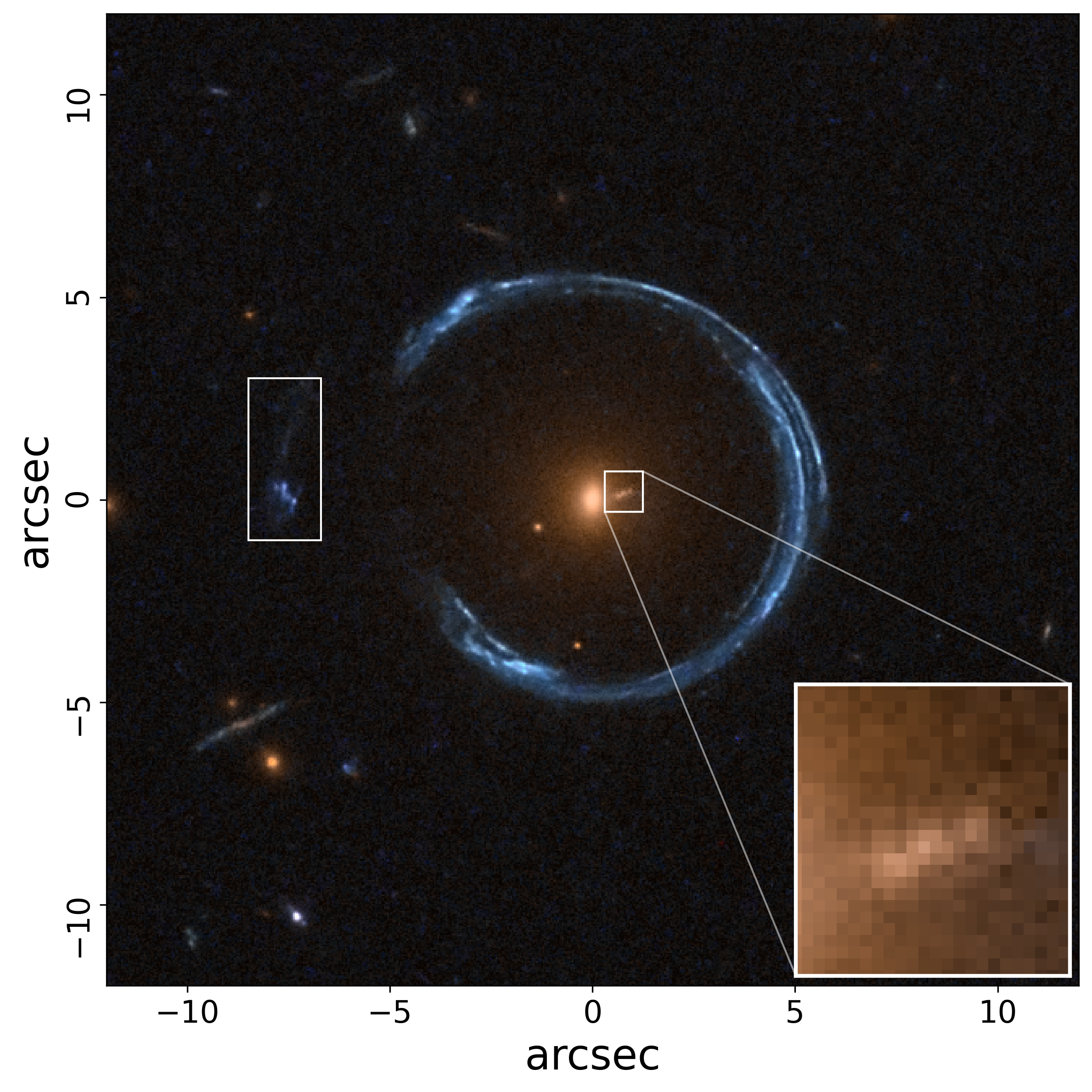}
    \caption{HST/WFC3 colour composite image of the Cosmic Horseshoe, created using the F814W, F606W, and F475W filters. The system is composed by the main deflector ($z_l = 0.44$); the eponymous Einstein ring of the Cosmic Horseshoe ($z_{\text{s2}} = 2.381$); and the radial arc and its counter-image ($z_{s1} = 1.961$), both highlighted. The inset shows the radial arc. The figure is oriented such that  north is up and east is left.}
    \label{fig:horseshoe_rgb}
\end{figure}

The Cosmic Horseshoe (SDSS J1148+1930) system was first identified by \cite{Belokurov2007} as part of the Sloan Digital Sky Survey (SDSS). The primary lens is a massive ETG at $z_l = 0.44$, with an estimated enclosed mass within the Einstein radius ($R_\text{Ein}$) of $\sim 5 \times 10^{12}M_\odot$ \citep[][]{Dye2008,Schuldt2019}. The radial arc and its counter-image correspond to a source at redshift $z_{s1} = 1.961$ (s1, hereafter), while the tangential arc is a star-forming galaxy \citep[][]{James2018} at redshift $z_{s2} = 2.381$ (s2, hereafter). In Fig. \ref{fig:horseshoe_rgb}, the radial arc and its counter-image are highlighted within white boxes, with a zoomed-in view of the radial image displayed in the inset.

\subsection{HST imaging}
The HST images used in this work were obtained with the Wide Field Camera 3 (WFC3) and downloaded from the Hubble Legacy Archive\footnote{\url{https://hla.stsci.edu/}} (HLA). Observations with the F475W, F606W, F814W, F110W, and F160W filters were conducted in May 2010\footnote{PropID: 11602, PI: Sahar Allam}, while the F275W filter data were taken in November 2011\footnote{PropID: 12266, PI: Anna Quider}. The data reduction followed the HLA pipeline, which employs the DrizzlePac\footnote{\url{https://drizzlepac.readthedocs.io/en/latest/index.html}} to process the images. This includes the combination of multiple exposures, correction for geometric distortion, subtraction of the sky background, and removal of cosmic rays. The final science images for the UVIS filters (F275W, F475W, F606W, and F814W) have a pixel scale of $0.04\arcsec$, while the IR filters (F110W and F160W) provide images with a pixel scale of $0.13\arcsec$.

We made use of images in the F475W and F814W filters for our analysis. The F475W band was selected for lens modelling, as the radial arc appears bluer and more distinct from the main-lens in this filter. Conversely, the F814W band was used to trace the light distribution and stellar mass of the primary deflector (see Section \ref{sec:methods}), as the main-lens is brighter in this filter and the radial arc is not visible, minimising contamination from s1. Both images were aligned using the \texttt{Astroalign} software \citep[][]{Beroiz}.

To construct the point spread function (PSF) for each filter, we identified two non-saturated stars from the Gaia DR2 catalogue \citep[][]{GaiaDR2} and performed a interactive PSF modelling using \texttt{PSFr} \citep[see, e.g.,][]{Birrer2019}. We follow the same procedure as \citet{Birrer2019}, applying a $90^\circ$ rotation symmetry based on the HST optical symmetry. This allows each star to be rotated four times, enabling PSF model estimation using a total of eight stacked images. The noise map for each pixel $i$ was calculated by combining the background noise $\sigma_{\text{bkgd}}$ and Poisson noise in quadrature: $\sigma^2_{\text{rms,i}} = \sigma^2_{\text{bkgd}} + \sigma^2_{\text{Poisson,i}}$. The background level was estimated as a constant value, measured from an empty region of sky near the main deflector, using the \texttt{astropy} sigma-clipping method. The Poisson noise was calculated from the effective exposure map and the intensity counts for each pixel.

\subsection{MUSE observations and kinematical map}

The integral-field spectroscopic observations were conducted using the VLT/MUSE instrument across three separate visits\footnote{ProgID: 094.B-0771, PI: Bethan James} and retrieved from the ESO Science Archive Facility\footnote{\url{https://archive.eso.org/scienceportal/home}}. The data covers a spectral range of  $4650-9300$\AA, sampled at  $1.25$\AA/px, with a mean spectral resolution of $\sim2.6$\AA \, at full width at half-maximum (FWHM; $\sigma \sim 50$\,km\,s$^{-1}$). The spatial pixel scale is $0.2\arcsec$, and the seeing during observations was $0.8\arcsec$. Data reduction followed the ESO Phase 3 Data Release, utilising the MUSE pipeline \citep[][]{Weilbacher2016}. The MUSE data cube was aligned with HST images by generating a collapsed image from the cube and using \texttt{Astroalign} to register it with the F475W HST observation.

To extract stellar kinematics from the MUSE data cube, we selected all pixels with a signal-to-noise ratio (SNR) greater than $2.5$, excluding regions exhibiting emission from s2. Pixels corresponding to the radial source position were also inspected, but no contribution from s1 was identified in the spectra. The remaining pixels were spatially binned using the Voronoi binning method of \citet{Cappellari2003} to achieve a minimum SNR of $15$. The SNR was calculated as the ratio of the average signal to the average noise in the rest-frame spectral range $5600-7600$\,\AA. Using the continuum rest-frame range $6000-6200$\,\AA\,, produced negligible differences in the results.

The mean velocity ($v$) and velocity dispersion ($\sigma_v$) in each Voronoi bin were measured using the penalized pixel fitting method, as implemented in the \texttt{pPXF} software \citep[][]{Cappellari2004}. For templates, we adopted the full X-shooter Spectral Library SSP models \citep[XLS-SSP-DR3;][]{Verro2022}, which offer a resolution of $\sim 13$ km\,s$^{-1}$ and a wavelength coverage from $3500$ to $24800$\,\AA. The XLS-SSP library was selected due to its high resolution, enabling convolution with the MUSE instrumental resolution after de-shifting the galaxy spectra. Each Voronoi bin was fitted over the wavelength range $5600-7600$\,\AA \, (galaxy-frame), with emission lines within this range masked. 

Uncertainties for $v$ and $\sigma_v$ were determined through Monte Carlo perturbations of the best-fit model, performing $200$ realisations and taking the standard deviation as the uncertainty. {Template selection is a known source of systematic uncertainty. Recently, \citet{Knabel2025} highlighted that template mismatch can dominate the error budget in high-precision applications, for instance in time-delay cosmography. However, for SMBH mass measurements, we consider its impact to be secondary relative to other sources of uncertainty, such as assumptions in the dynamical modelling (e.g., anisotropy profile and intrinsic shape). A detailed investigation of template choice effects on SMBH mass measurements is beyond the scope of this work.}

We computed the velocity second moment as $v_\text{rms} = \sqrt{v^2 + \sigma_v^2}$, finding that rotational velocities are negligible, indicating the galaxy is dominated by the velocity dispersion.

\begin{figure*}
    \centering
    \includegraphics[width=0.90\textwidth]{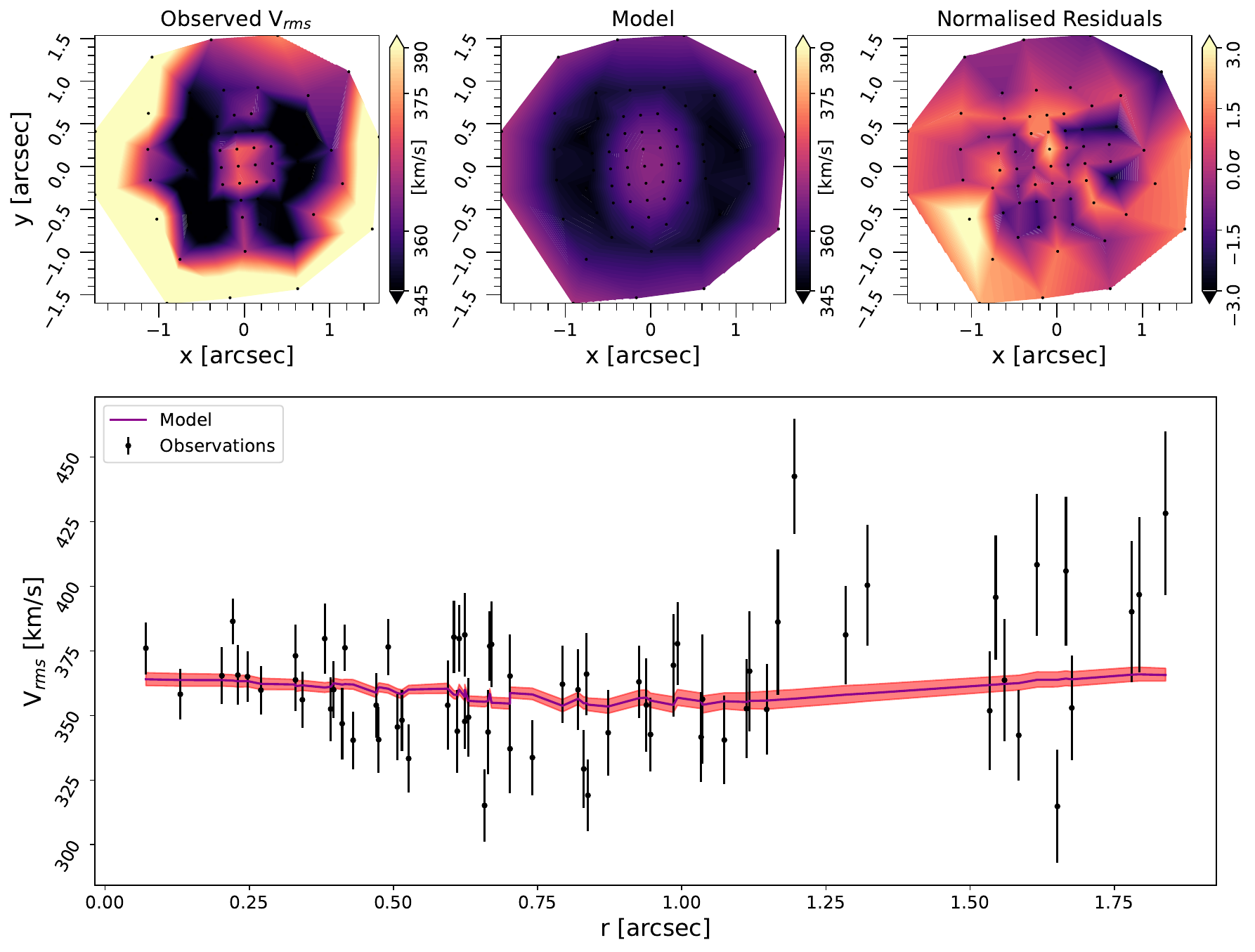}
    \caption{Stellar dynamics fiducial model. The top panels show the observed $v_\text{rms}$ kinematic map (left), the median kinematic model (centre), and the normalised residuals (right). The bottom panel presents the radial kinematic profile (black dots) alongside the median model and its $1\sigma$ credible region. The black dots in the top panels mark the centroids of the Voronoi bins.}
    \label{fig:fiducial_dyn_model}
\end{figure*}

The observed $v_{\rm rms}$ map and radial profile are shown in Fig.~\ref{fig:fiducial_dyn_model}. The profile is nearly flat in the galaxy's central regions ($<0.5\arcsec$) but increases for $r > 0.75\arcsec$. This rising behaviour at larger radii was previously reported by \citet{Spiniello2011} using long-slit spectroscopy of the Cosmic Horseshoe and is a common feature among BCGs \citep[e.g.,][]{Newman2015,Smith2017}. Notably, the outermost bins exhibit larger error bars, reflecting the low SNR in these regions.

The effective velocity dispersion of the main-lens was determined by co-adding all spectra within the galaxy's effective radius ($R_e = 2.10\arcsec$)\footnote{The effective radius was determined using the galaxy's MGE surface brightness model. See Section \ref{sec:MGE} for more details.} and fitting the integrated spectrum with \texttt{pPXF}, as outlined earlier. This analysis yielded  $\sigma_e = 366 \pm 6$\,km\,s$^{-1}$. We confirmed that using the flux-weighted method present in \citet{Emsellem2007} lead to a similar result.

\section{Methods}\label{sec:methods}
We construct a fully self-consistent mass model for the main lensing galaxy by jointly modelling its SGL effect and spatially-resolved stellar velocity dispersion. This combined approach has been successfully applied in previous studies to constrain the mass profiles of ETGs in tests of modified gravity \citep[e.g.,][]{Collett2018, Melo-Carneiro2023} and to investigate the distribution of baryonic and DM content within galaxies \citep[e.g.,][]{Barnabe2012, Wang2022}. By simultaneously leveraging both methods, we mitigate degeneracies that arise when using them independently, such as the mass-anisotropy degeneracy \citep[e.g.,][]{Gerhard1993} and the mass-sheet degeneracy \citep[e.g.,][]{Gorenstein1988}. {Furthermore, the Cosmic Horseshoe system is a compound lens with two sources at different redshifts \citep[][]{Schuldt2019}. The joint modeling of these sources can also help break (or at least alleviate) these degeneracies present in the lens model} \citep[e.g.,][]{Enzi2024}.

\subsection{Gravitational Lensing}\label{sec:Grav_lensing}
\subsubsection{Formalism}
For a single-source plane lens configuration, the source plane position, $\bmath{\beta}$, relates to the observed lensed position, $\bmath{\theta}$, via the lens equation:

\begin{equation}\label{eq:lens_equation}
    \bmath{\beta} = \bmath{\theta} - \bmath{\alpha}(\bmath{\theta}),
\end{equation}
where $\bmath{\alpha}$ is the reduced deflection angle, given by:

\begin{equation}\label{eq:reduced_angle}
    \bmath{\alpha}(\bmath{\theta}) = \frac{1}{\pi} \int d^2\bmath{\theta}^\prime \kappa(\bmath{\theta}^\prime) \frac{\left(\bmath{\theta} - \bmath{\theta}^\prime \right)}{\,\,\left| \bmath{\theta} - \bmath{\theta}^\prime\right|^2}.
\end{equation}
Here, the convergence $\kappa = \Sigma / \Sigma_\text{crit}$, represents the surface mass density of the lens scaled by the critical surface density, $\Sigma_\text{crit}$,  defined as:

\begin{equation}
    \Sigma_\text{crit} = \frac{c^2}{4\pi G}\frac{D_s}{D_l D_{ls}},
\end{equation}
where $D_{ls}$, $D_l$ and $D_s$ are the angular diameter distances between the lens and source, lens and observer, and source and observer, respectively.

One can notice that changing the source redshift, only the angular diameter distances are changed in the reduced deflection angle, Eq. \ref{eq:reduced_angle}. For two light rays passing through the same point in the lens plane, but originating from different source planes, the relationship between their deflection angles can be expressed as a scaling factor, $\eta$, determined by the ratio:

\begin{equation}\label{eq:beta_factor}
    \frac{\bmath{\alpha}_1}{\bmath{\alpha}_2} = \frac{D_{ls1}}{D_{s1}}\frac{D_{s2}}{D_{ls2}} \equiv \eta, \quad \text{with $z_{s2} > z_{s1}$},
\end{equation}
and where the subscripts $s1$ and $s2$ refers to the sources s1 and s2.

In cases where multiple sources are aligned along the line-of-sight (LOS), the gravitational field of the first source lenses the light from the second source before it is further deflected by the main-lens. To account for these effects, a full multiple-lens-plane formalism is required  \citep[][]{Schneider1992}.  The single-plane lens equation (Eq. \ref{eq:lens_equation}) can then be generalised into the compound lens equation:

\begin{equation}\label{eq:multiplane}
    \bmath{\theta}_k = \bmath{\theta}_1 - \sum_{j=1}^{k-1}\eta_{jk} \bmath{\alpha}_j\left(\bmath{\theta}_j\right),
\end{equation}
where $\bmath{\theta}_k$ is the angular position of a light ray in the $k^\text{th}$ plane, and $\bmath{\alpha}_j$ is the angular deflection caused by the $j^{\text{th}}$ lens acting on rays originating from the furthest redshift source plane.  The factor $\eta_{jk}$ is the scaling factor as defined in Eq. \ref{eq:beta_factor}, i.e.,

\begin{equation}
    \eta_{jk} = \frac{D_{jk}}{D_k}\frac{D_{s}}{D_{js}},
\end{equation}
with $s$ being the most distant source. In the case of just one lens and one source, Eq. \ref{eq:multiplane} reduces to Eq. \ref{eq:lens_equation}, with $\bmath{\theta}_2 = \bmath{\beta}$, $\bmath{\theta}_1 = \bmath{\theta}$, and $\eta_{jk} = 1$.

To reconstruct the unlensed sources, the lens model must, in principle, account for deflections from both the main deflector and source s1 on s2. However, in the Cosmic Horseshoe system, the deflection contribution from s1 to the total deflection angle experienced by s2 is expected to be negligible for three main reasons: (i) the sources lie at similar redshifts, making s1 an inefficient lens; (ii) the positions of the radial image and its counter-image indicates that s1 is not closely aligned with s2, and the lensed images of s2 do not pass near s1; and (iii) the low observed lensed surface brightness of s1, combined with its small size after reconstruction (see Section \ref{sec:results}), suggests that it is a low-mass galaxy. {To further support this assumption, we perform a compound lens model including s1’s mass contribution, which confirms its negligible impact.}

\subsubsection{Lens modelling}\label{sub:lens_modelling}
For the lens modelling, we employ the open-source software \texttt{PyAutoLens} \citep[][]{Nightingale2018,PyAutoLens}, which implements a Bayesian version \citep[][]{Suyu2006} of the semi-linear inversion (SLI) method \citep[][]{Warren2003}. For a given set of non-linear parameters (describing the lens mass model and/or the source), the code linearly ray-traces image-pixels from the image plane back to the source plane, reconstructing the source emission using an adaptive mesh grid\footnote{This assumes that the lens light was subtracted already.}.

In \texttt{PyAutoLens} the {\it native} likelihood is that described in \citet[][eq. 19]{Suyu2006} and \citet[][eq. 5]{Dye2008}. This approach incorporates the instrumental PSF blurring alongside regularisation terms for the pixelised source reconstruction, which helps to mitigate ill-posed solutions during the linear inversion. We refer the readers to these references for more details \footnote{Those interested in the technical aspects and implementation within \texttt{PyAutoLens} may also check the following notebooks: \url{https://github.com/Jammy2211/autolens_workspace/tree/main/notebooks/advanced/log_likelihood_function}}.

To remove over- and under-magnified solutions \citep[][]{Maresca2021}, we trace two image plane regions of s1 back to the source plane, where they are expected to overlap. If these conjugate regions fail to overlap after delensing, we penalise the likelihood of this solution by a factor proportional to the distance between the two regions in the source plane. This approach effectively removes unphysical solutions resembling over/under-magnified versions of the data, without the need for parameter fine-tuning. Using regions, rather than pairs of conjugate points, also mitigates the risk of selecting incorrect pairs. This is particularly relevant in the case of the Cosmic Horseshoe system, where the radial image is very faint and embedded within the lens light, making it challenging to reliably identify conjugate pairs between the radial image and its counter-image. Further details of this approach are provided in Appendix \ref{ap:conjugated}.

In this work, our primary goal is to constrain the inner mass distribution of the main deflector through joint modelling of the radial image of s1 and its counter-image, along with the host galaxy’s stellar kinematics. As a result, we do not attempt to model the full Einstein ring of s2 in our main analysis. Nonetheless, the lensed emission from s2 traces the total projected mass distribution on larger scales and thus provides valuable complementary constraints on the global mass profile of the main deflector.

\begin{figure*}
    \centering
    \subfloat{
    \includegraphics[width=0.65\columnwidth]{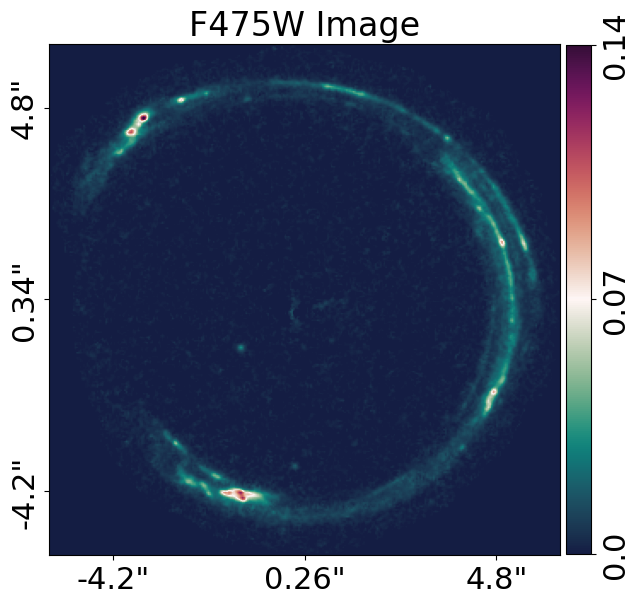}
    }
    \quad
    \subfloat{
    \includegraphics[width=0.65\columnwidth]{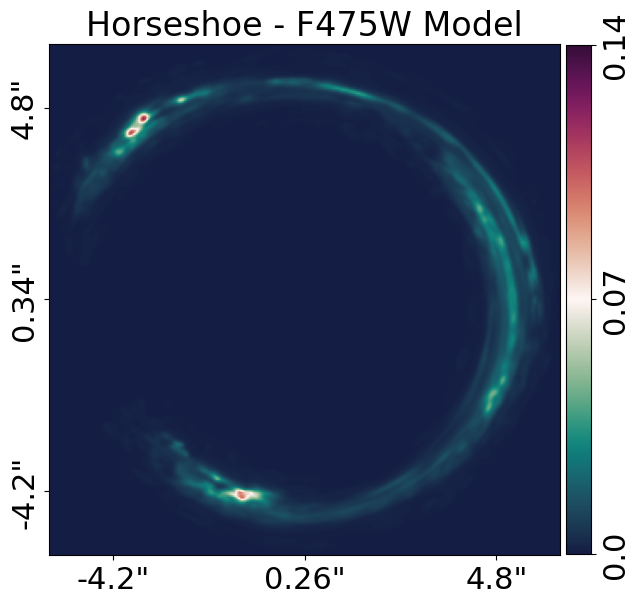}
    }
    \quad
    \subfloat{
    \includegraphics[width=0.65\columnwidth]{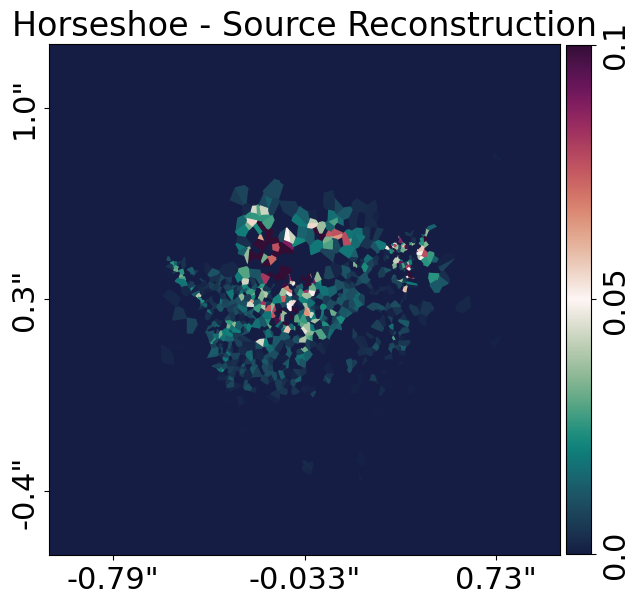}
    }
   \caption{Fits to the Cosmic Horseshoe in the F475W filter. From left to right, the panels display the lens-subtracted image, the highest-likelihood EPL model, and the reconstructed source s2 at $z=2.381$. To enhance the efficiency of the lens modelling, we applied a mask around the lensed source and only modelled pixels within the masked region, as shown in the central panel. All images are in units of electrons per second. Additional details about the lens modelling are provided in Appendix \ref{ap:EPL}.}
    \label{fig:Horseshoe_EPL}
\end{figure*}

To incorporate this information, we first modeled the lensed emission from source s2 independently, using an elliptical power-law (EPL) mass profile for the main deflector implemented in \texttt{PyAutoLens}. This fit yields a mass within the Einstein radius of $M_{\text{Ein}} = 5.46 \times 10^{12}\,M_{\sun}$, corresponding to an effective Einstein radius of $R_\text{Ein}=5.08\arcsec$. These values agree with those reported by \citet{Dye2008} and \citet{Schuldt2019}. We then incorporate this result into our combined lensing+dynamics analysis by imposing a Gaussian prior on  $M_{\text{Ein}}$ with mean $5.46 \times 10^{12} M_{\sun}$ and standard deviation of $\sigma_{M_{\text{Ein}}} = 0.27 \times 10^{12} M_{\sun}$. Figure~\ref{fig:Horseshoe_EPL} shows the image pixels used for the EPL fit, the highest-likelihood model, and the reconstructed source. Further details of the EPL fitting procedure are provided in Appendix~\ref{ap:EPL}.

To test the assumption that compound lensing by the first source is  negligible, we also ran a compound lens model with the \texttt{HercuLens} code \citep[][]{Galan2022, Enzi2024}, including a singular isothermal
ellipse  mass component for s1. For this compound model the Einstein radius of the primary lens never changed by more than 0.6\%, which is substantially smaller than the width of the Gaussian we have adopted for the mass of the primary lens within the main Horseshoe.

While this prior provides an additional constraint on the large-scale mass distribution, we test its impact in Section~\ref{sec:results} and find that excluding it from the likelihood has no effect on the inferred SMBH mass.

\subsection{Dynamical modelling}

We described the dynamical state of the system using the Jeans formalism for a steady-state axisymmetric configuration \citep[][]{Binney_Tremaine2008}. In spherical coordinates $(r, \theta, \phi)$, and assuming the velocity ellipsoid is aligned with the spherical coordinate system, the Jeans equations are \citep[][]{Cappellari2020}:

\begin{equation}\label{eq:Jeans}
\begin{aligned}
    \frac{\partial (\nu \overline{v^2_r})}{\partial r} 
    + 
    \frac{2\nu\overline{v^2_r} -\nu\overline{v^2_\theta} -\nu\overline{v^2_\phi} }{r}
    &=
    -\nu\frac{\partial \Phi}{\partial r}, \\
    \frac{\partial (\nu \overline{v^2_\theta})}{\partial \theta} 
    +
    \frac{\nu \overline{v^2_\theta} - \nu\overline{v^2_\phi}}{\tan{\theta}}
    &=
    -\nu\frac{\partial \Phi}{\partial \theta},
\end{aligned}
\end{equation}
where $\Phi$ is the total gravitational potential, $(v_r, v_\theta, v_\phi)$ the velocities in spherical coordinates, and  $\nu$ is the intrinsic luminosity density.

Defining the stellar anisotropy as 
\begin{equation}
    \beta_\text{star} = 1 - \frac{{\overline{v^2_\theta}} } {{\overline{v^2_r}}} \equiv 1 - \frac{{\sigma^2_\theta}} {{\sigma^2_r}},
\end{equation}
and applying the boundary condition $\nu \overline{v^2_r} = 0$ as $r \rightarrow 0$, the Jeans Eqs. \ref{eq:Jeans} simplify to: 

\begin{equation}\label{eq:Jeans_solution1}
    \nu \overline{v^2_\phi} 
    = 
    \left(1 - \beta_\text{star}\right)\left[   \nu\overline{v^2_r} + \frac{\partial \left( \nu \overline{v^2_r}\right)}{\partial \theta}\tan{\theta}   \right] 
    +
    \nu \frac{\partial \Phi}{\partial \theta}\tan{\theta},
\end{equation}
and
\begin{equation}\label{eq:Jeans_solution2}
    \nu \overline{v^2_r} 
    =
    {\displaystyle \int_{r}}^\infty \left( \frac{r^\prime}{r} \right)^{2\beta_\text{star}}\Psi(r^\prime, \theta^\prime)dr^\prime,
\end{equation}

where

\begin{equation}
    \begin{aligned}
        &\theta^\prime = \arcsin{\left[   \left( \frac{r^\prime}{r} \right)^{\beta_\text{star} - 1}\sin{\theta}    \right]}, \quad \text{and}\\
        &\Psi(r, \theta) = \nu \left(    \frac{\partial \Phi}{\partial r} -\frac{\tan{\theta}}{dr}\frac{\partial \Phi}{\partial \theta}    \right).
    \end{aligned}
\end{equation}
By solving Eqs. \ref{eq:Jeans_solution1} and \ref{eq:Jeans_solution2}, the velocity moments can be integrated along the LOS \citep[see, e.g., Section 3 on][]{Cappellari2020} to compute the observables, which can be compared to the galaxy's observed $v_\text{rms}$ map.

For this purpose, we employed the Jeans Anisotropic Modelling (JAM) method \citep[][]{Cappellari2008,Cappellari2020}, as implemented in the \texttt{Jampy} software, to perform the stellar dynamical modelling.  This approach assumes that the galaxy's mass distribution can be parameterised as a sum of concentric elliptical Gaussians (see Section \ref{sec:MGE}). 

For the dynamical modelling, we evaluated the goodness-of-fit using a $\chi^2$-likelihood, which compares the observed data to the model predictions convolved with the MUSE PSF\footnote{Here assumed to be the observed seeing.}. This ensures that the effects of seeing are properly accounted for in the analysis.

\subsection{Multi-Gaussian Expansion (MGE)}\label{sec:MGE}
The stellar light and mass component are modelled using the MGE method \citep{Emsellem1994E,Cappellari2002}, which parametrises the stellar surface brightness as a sum of concentrict elliptical Gaussians with the same orientation.  If $I(x^{\prime},y^{\prime})$ represents the stellar surface brightness, its MGE parametrisation is given by

\begin{equation}\label{eq:surf_MGE}
    I(x^{\prime},y^{\prime}) = \sum_{j=1}^{N} \frac{L_j}{2\pi \sigma_{j}^{2} q^{\prime}_j}\exp{\left[-\frac{1}{2\sigma_{j}^{2}} \left(x^{\prime 2} + \frac{y^{\prime 2}}{q_{j}^{\prime 2}}\right)\right]},
\end{equation}
where $N$ is the total number of Gaussians. The $j^{\text{th}}$ Gaussian component has a total luminosity $L_j$, an observed projected axial ratio $0 \leq q^{\prime}_j  \leq 1$, and a dispersion $\sigma_j$ along the semi-major axis, aligned with $x^{\prime}$-axis.

The three-dimensional luminosity density $\nu$ (and corresponding mass density) is obtained by deprojecting Eq. \ref{eq:surf_MGE}, assuming an inclination angle $i$, where $i = 90^{\circ}$ corresponds to an edge-on orientation. The luminosity density $\nu$ is then converted into stellar mass density using a mass-to-light ratio, $\Upsilon_\star$. 

Assuming an oblate axisymmetric model, the stellar mass density profile in cylindrical coordinates $(R, \phi, z)$ is given by \citep[][]{Cappellari2002}:

\begin{equation}\label{eq:MGE mass density}
    \rho(R, z) = \sum_{j=1}^N \frac{M_j}{(2\pi)^{3/2} \sigma_{j}^3 q_j} \exp{\left[-\frac{1}{2 \sigma_j^2} \left(R^2 + \frac{z^2}{q_{j}^2}\right)\right]}.
\end{equation}
Here, $M_j = \Upsilon_\star L_j$ represents the mass of the $j^{\text{th}}$ Gaussian component, with $L_j$ luminosity and $\sigma_j$ dispersion. The deprojected axial ratio, $q_j$, is given by
\begin{equation}\label{eq:_q_deproj}
    q_{j}^2 = \frac{q_{j}^{\prime 2} - \cos^2{i}}{\sin^2{i}}.
\end{equation}

\subsection{Mass profile}
We described the mass profile of the main deflector using the MGE method, assuming a multicomponent mass model composed by (i) a stellar mass component; (ii) a DM halo component;  and (iii) an additional central mass concentration representing an SMBH. 

The stellar component was derived by deprojecting the observed surface brightness profile, as by Eq. \ref{eq:surf_MGE}. The DM halo follows a generalised Navarro-Frank-White \citep[gNFW;][]{Wyithe2001} profile,

\begin{equation}\label{eq:gNFW}
    \rho(r) = \rho_s  \left( \frac{r}{r_s} \right)^{-\gamma_{\text{DM}}} \left(1 + \frac{r}{r_s}\right)^{\gamma_{\text{DM}} - 3},
\end{equation}
where $\rho_s$ is a characteristic density at the scale radius $r_s$, and $\gamma_{\text{DM}}$ is the inner density slope that allows the profile to be cuspier ($\gamma_{\text{DM}} > 1$) or cored ($\gamma_{\text{DM}} = 0$). For $\gamma_{\text{DM}} = 1$, the profile reduces to the classical NFW \citep[][]{Navarro1997}. To include this halo contribution in the dynamical model, we followed \citet{Cappellari2013} and also parametrised the DM component using the MGE method\footnote{This parametrisation is performed iteratively, starting with 20 Gaussian components and adding more if the absolute deviation from the target analytical profile exceeds 10\% in any radial bin. The process continues until all deviations fall below the threshold or a maximum of 45 components is reached. Based on our tests, 20 Gaussians are typically sufficient to accurately represent NFW or gNFW profiles.}. Finally, the SMBH is modelled as an additional Gaussian with a small scale radius ($\sigma = 0.01\arcsec$).

The total mass distribution of the main-lens galaxy is then described by Eq. \ref{eq:MGE mass density}, with the summation over $N_\text{star} + N_\text{DM} + N_\text{BH}$ Gaussians, where $N_j$ represents the number of Gaussians used to parameterise each respective mass component.

We also incorporate an external lensing shear to account for the perturbations of structures near the LOS, besides s1. The external shear field is parameterised by the two elliptical components $(\epsilon^{\text{sh}}_1, \epsilon^{\text{sh}}_2)$. From these components, the shear magnitude $\gamma^{\text{sh}}$ and shear angle $\phi^{\text{sh}}$, measured counter-clockwise from north, are obtained as:

\begin{equation}\label{eq:shear}
    \gamma^{\text{sh}} = \sqrt{{\epsilon^{\text{sh}}_1}^2 + {\epsilon^{\text{sh}}_2}^2}, \quad \tan{\left(2\phi^{\text{sh}}\right)} = \frac{\epsilon^{\text{sh}}_2}{\epsilon^{\text{sh}}_1}.
\end{equation}

\subsection{Joint modelling}\label{sec:joint_modelling}
Since lensing and dynamical data are independent, the joint likelihood is constructed as the product of their individual likelihood functions. In addition to these terms, the likelihood includes two additional terms: one enforcing consistency with the mass enclosed within the Einstein radius, as defined by the lensed image of s2 ($M_{\text{Ein}}$), and the penalty term defined in Sec. \ref{sub:lens_modelling} that punishes solutions where s1 is significantly over- or under-magnified.

We break the modelling process into the following stages, so as to make it more tractable

\begin{enumerate}
    \item Dynamical Model Fitting: We begin by fitting only the dynamical model, but with the addition of the prior term for \,$M_{\text{Ein}}$ and the conjugate regions of s1 image. This provides an initial estimate of the mass parameters.

    \item Source Grid Model of s1: Using the highest-likelihood dynamical model, we then sample the pixelised source grid of s1. A Voronoi grid is used to allocate more pixels to highly magnified regions of the source plane, along with a constant regularisation term. In \texttt{PyAutoLens}, this is represented by an \texttt{Overlay} mesh grid, with \texttt{VoronoiNN} pixelisation, and \texttt{ConstantSplit} regularisation.

    \item Mass Parameter Resampling: After obtaining the source grid, we resample the mass parameters using the full (lensing+dynamics) likelihood, with s1 reconstructed based on the highest-likelihood model from the previous chain. This step improves the estimate of the mass parameters.

    \item Source Plane Resampling: Using the highest-likelihood model of the previous stage, we resample the parameters describing the source plane pixelisation and regularisation. The source is now reconstructed in the source plane using a brightness-adaptive Voronoi mesh grid, with Natural Neighbour interpolation \citep[][]{Sibson1981} and a regularisation scheme that adapts the degree of smoothing based on the source's surface brightness. This ensures that brighter regions are reconstructed at higher resolution, while regions with lower SNR are more regularised. In \texttt{PyAutoLens} notation, this corresponds to a \texttt{KMeans} mesh grid, with \texttt{VoronoiNN} pixelisation, and \texttt{AdaptiveBrightnessSplit} regularisation.

    \item Final Resampling of the Lens Macro Model: Finally, using the source grid that adapts to the source morphology, we sample the lens macro model once more, keeping the hyper-parameters of the source fixed at the highest-likelihood result of the previous step.
    
\end{enumerate}
Throughout all these stages, the same priors are applied across all chains for each given lens mass macro model. Further details are provided in Appendix \ref{ap:priors}.

We sampled for the posterior distribution of the non-linear parameters using the nested sampler \texttt{dynesty} \citep[][]{Speagle2020}, a Python implementation of the nested sampling algorithm \citep[][]{Skilling2006} designed for estimating Bayesian posteriors and evidences. The Bayesian evidence, $\mathcal{Z}$, which is the integral of the likelihood times the prior over the entire multidimensional parameter space, can be used to rank different models for the same dataset, providing an objective way to compare models \citep[e.g.,][]{Liddle2006}. 

The Bayesian evidence naturally incorporates a penalty for increased model complexity – penalising models with additional free parameters-, and provides an objective basis for the principle of Occam’s razor. Typically, when comparing models using the difference in their log-evidences, $\Delta \ln\mathcal{Z}$, a difference in $\Delta \ln{\mathcal{Z}} > 1$ is considered significant, $ >2.5$ strong, and $>5$ is decisive, in favour of the model with the highest evidence \citep[e.g.][]{Jeffreys1961}. 

\section{Results}\label{sec:results}

To disentangle the baryonic matter from the DM contribution and to reveal the radial arc, a model for the lens light distribution is essential. We fitted the lens light in both HST/WFC3 filters, F814W and F475W, using the MGE method described in Section \ref{sec:MGE}. During the fitting process, the radial and tangential lensed images were masked to minimise contamination from the source planes. The PSF effects were incorporated by modelling the \texttt{PSFr} model as a sum of circular Gaussians. The F814W lens light model was used to trace the stellar distribution, as it is a better tracer of the stellar mass budged and has less contamination from the radial arc. For the lens modelling, we utilised the F475W image after the lens light subtraction, as the radial source is better seem in this band. The left panel of Fig. \ref{fig:Horseshoe_EPL} shows the F475W lens-subtracted image, and Table \ref{table:MGE} presents the MGE decomposition for the F814W band.

In the following sections, we discuss our fiducial model and explore various mass models to assess potential systematic effects on the SMBH mass inference. Table \ref{tab:mass_models} summarises these models and their resulting $M_\text{BH}$ values. Unless otherwise, all parameter estimates are derived from the final sampling chain, with reported values representing the median of each parameter's one-dimensional marginalised posterior distribution, with uncertainties corresponding to the $16^\text{th}$ and $84^\text{th}$ percentiles.

\begin{table}
\centering
\caption{MGE components of the HST/F814W image. The columns, from left to right, show the surface brightness of each Gaussian component, the MGE width, and the observed axis ratio. The MGE units were converted from counts to physical values following \citet{Trick2016}, accounting for the redshift dimming effect and assuming a solar Space Telescope magnitude (STmag) of $M_{\sun,\text{F814W}} = 5.35$ from \citet{Willmer_2018}.}
\label{table:MGE}
\begin{tabular}{lcc} 
\hline
 \makecell{$I$  \\ $[$L$_\odot$/pc$^2]$}  & \makecell{$\sigma$ \\ $[$arcsec$]$}& \makecell{$q^\prime$ \\ } \\
\hline
2416.68  &      0.0198   &   0.825  \\
5883.51  &      0.1379   &   0.650  \\
351.98	 &      0.2026   &   1.000  \\
1630.70  &      0.2161   &   0.650  \\
770.51	 &      0.3609   &   0.673  \\
425.76   &      0.5039   &   1.000  \\
171.75	 &      1.1015   &   1.000  \\
70.03    &      3.2367   &   1.000  \\ [1ex]
\hline
\end{tabular}
\end{table}

\subsection{Fiducial model} \label{results:fiducial}

The fiducial model ({\bf M1}) is composed by a stellar mass component, an elliptical NFW\footnote{This is obtained by taking $\gamma_\text{DM} = 1$ in Eq. \ref{eq:gNFW}.} halo, and a central SMBH. 

We assume that the stellar component follows the observed surface brightness distribution of the main-lens, scaled by a constant $\Upsilon_\star$. The DM component is assumed to be concentric with the baryonic matter and have the same alignment. Additionally, the scale radius $r_s$ is fixed at $10$ times the stellar effective radius, as seen in simulations \citep[e.g.,][]{Kravtsov2013} and used in other SGL studies \citep[e.g.,][]{Sonnenfeld2015}.

Beyond the mass parameters, the fiducial model includes four additional parameters: two elliptical components $(\epsilon^{\text{sh}}_1, \epsilon^{\text{sh}}_2)$ that describe the external lensing shear; a constant stellar anisotropy parameter, $\beta_\text{star}$; and the inclination angle, $i$, of the lens relative to the LOS. Therefore, the fiducial model has eight parameters: $\left(i, \beta_\text{star}, \Upsilon_\star, \log_{10}{\rho_s}, q_\text{DM}, \log_{10}{M_\text{BH}}, \epsilon^{\text{sh}}_1, \epsilon^{\text{sh}}_2\right)$. 

\begin{figure}
	\includegraphics[width=\columnwidth]{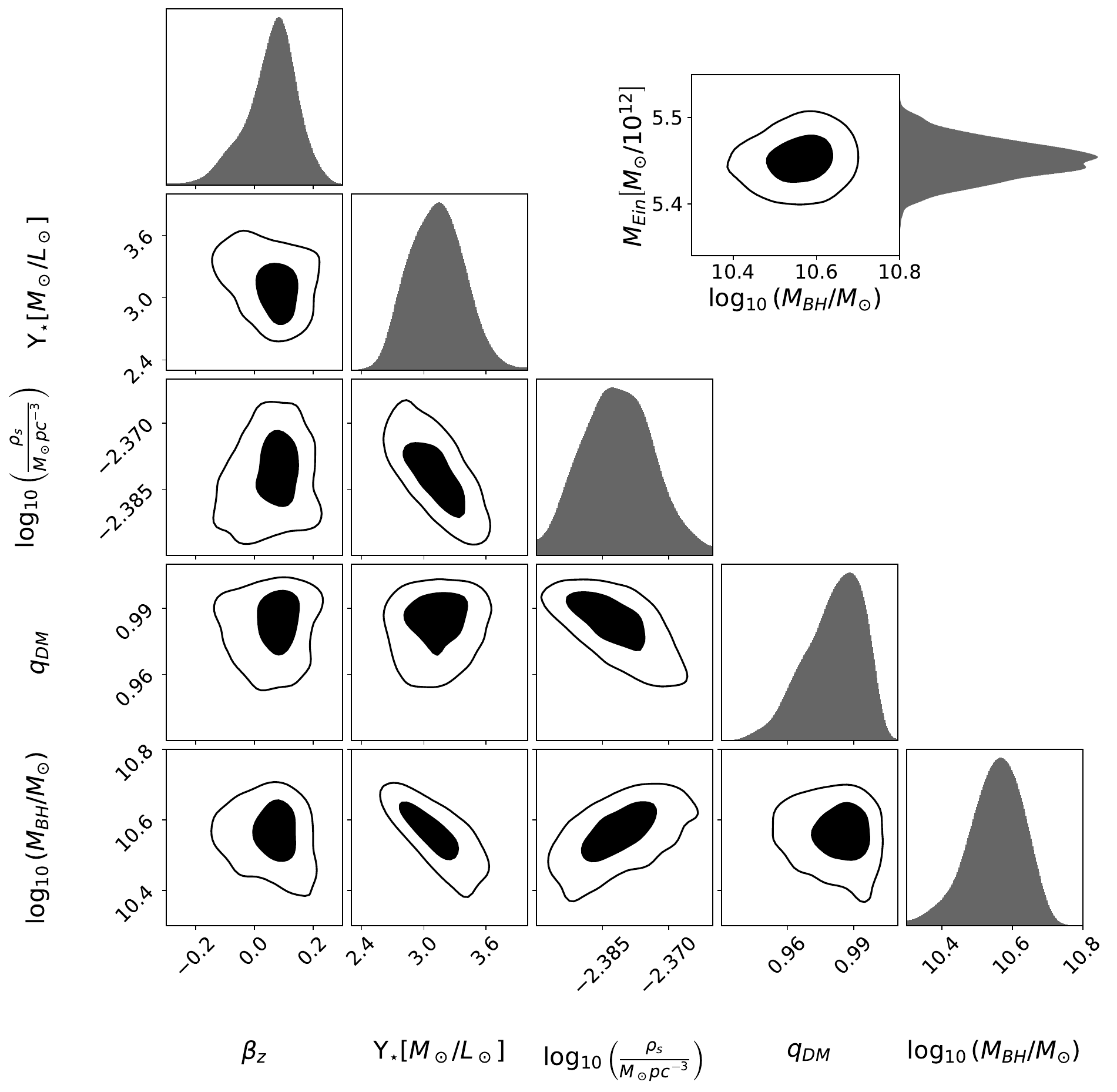}
    \caption{Two-dimensional posterior distributions for the parameters of the fiducial model. Only parameters that show a strong degeneracy with the SMBH mass are displayed. The inset plot on the top-right present the covariance between the SMBH mass and Einstein mass within the Cosmic Horseshoe ring. Contours are the $1$ and $2\sigma$ credible intervals, respectively.}
    \label{fig:corner_fiducial}
\end{figure}

The median of the one-dimensional marginalised posterior of the fiducial model and their uncertainties are summarised in Table \ref{table:fiducial_posterior}. Fig. \ref{fig:corner_fiducial} presents the two-dimensional posterior distributions for the SMBH mass and other parameters that exhibit significant degeneracies with it.

\begin{table}
\centering
\caption{Inferred median and $1\sigma$ credible intervals for the parameters of our fiducial
model.}
\renewcommand{\arraystretch}{1.5}
\label{table:fiducial_posterior}
\begin{tabular}{lc} 
\hline
 \makecell{Parameter \\ }  & \makecell{Posterior \\ (median with $1\sigma$ uncertainties)} \\
\hline
$i$ $[^{\circ}]$                    &      $65^{+15}_{-11}$           \\
$\beta_\text{star}$                           &      $0.07^{+0.06}_{-0.10}$     \\
$\Upsilon_\star$ $\left[M_{\sun}/L_{\sun}\right]$ &      $3.13^{+0.25}_{-0.26}$     \\
$\log_{10}\left(\frac{\rho_s}{M_\odot \text{pc}^{-3}}\right)$  &      $-2.38^{+0.01}_{-0.01}$     \\
$q_\text{DM}$                       &      $0.98^{+0.01}_{-0.02}$    \\
$\log_{10}(M_\text{BH}/M_{\sun})$   &      $10.56^{+0.07}_{-0.08}$     \\ 
$\epsilon^{\text{sh}}_1$            &      $-0.01^{+0.01}_{-0.01}$     \\
$\epsilon^{\text{sh}}_2$            &      $-0.05^{+0.01}_{-0.01}$     \\ 
$M_\text{Ein}\left[{M_{\sun}}/{10^{12}}\right]$ & $ 5.45^{+0.02}_{-0.03}$ \\ [1ex]
\hline
\end{tabular}
\end{table}

Our fiducial model favours an SMBH mass of $\log_{10}(M_\text{BH}/M_{\sun}) = 10.56^{+0.07}_{-0.08}$, therefore an UMBH.  The SMBH mass shows a  notable degeneracy with the mass-to-light ratio and the DM characteristic density, as shown in Fig. \ref{fig:corner_fiducial}. However, no significant degeneracy was observed between the SMBH mass and the anisotropy parameter. Similarly, the Einstein mass within the Cosmic Horseshoe, $M_\text{Ein}$, does not exhibit a strong degeneracy with $M_\text{BH}$ - top-right inset plot in Fig. \ref{fig:corner_fiducial}. 

The median stellar mass-to-light ratio, $\Upsilon_\star = 3.13^{+0.25}_{-0.26} M_{\sun}/L_{\sun}$, is in agreement with values inferred from stellar population synthesis models of other ETGs \citep[e.g.,][]{Conroy2012,Gu2022}. This is a reasonable value for a massive elliptical galaxy: the expected value is $\sim 4 M_{\sun}/L_{\sun}$ for a simple stellar population with an age of 10 Gyr and solar metallicity \citep[][]{Vazdekis2012,Ricciardelli2012,Verro2022}.

We recovered a total projected mass within $R_\text{Ein}$ of $M_\text{Ein} = 5.45^{+0.02}_{-0.03}\times 10^{12}M_{\sun}$, consistent with our prior and the EPL model, albeit slightly higher than the value reported by \citet{Dye2008} and \citet{Schuldt2019}. The DM fraction within $R_e$ was found to be $f_\text{DM}(\leq R_e) = 0.72^{+0.02}_{-0.02}$, which is also higher than \citet{Schuldt2019} and \citet{Spiniello2011}, though it remains consistent within the $1\sigma$ range.

\begin{figure*}
    \centering

    \begin{minipage}[t]{0.48\textwidth}
        \centering
        \subfloat{\includegraphics[width=0.8\columnwidth]{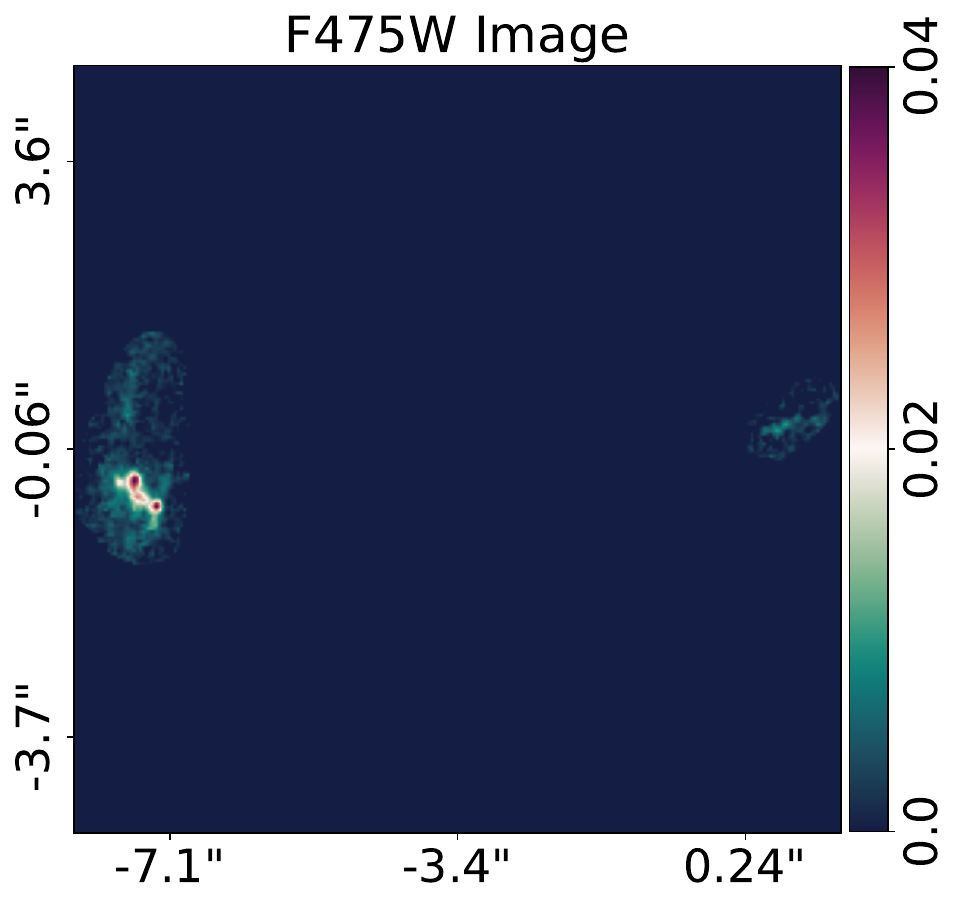}}
                
        \subfloat{\includegraphics[width=0.8\columnwidth]{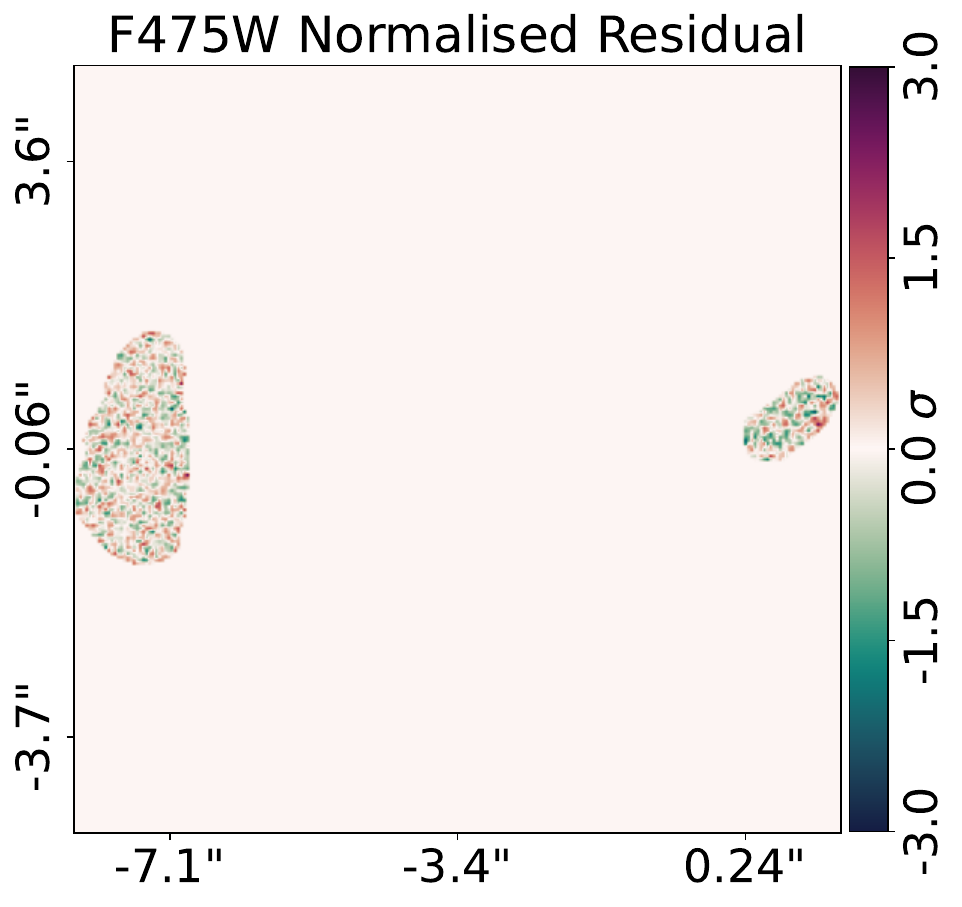}}
    \end{minipage}
    \hspace{-1cm}
    \begin{minipage}[t]{0.48\textwidth}
        \centering
        \subfloat{\includegraphics[width=0.8\columnwidth]{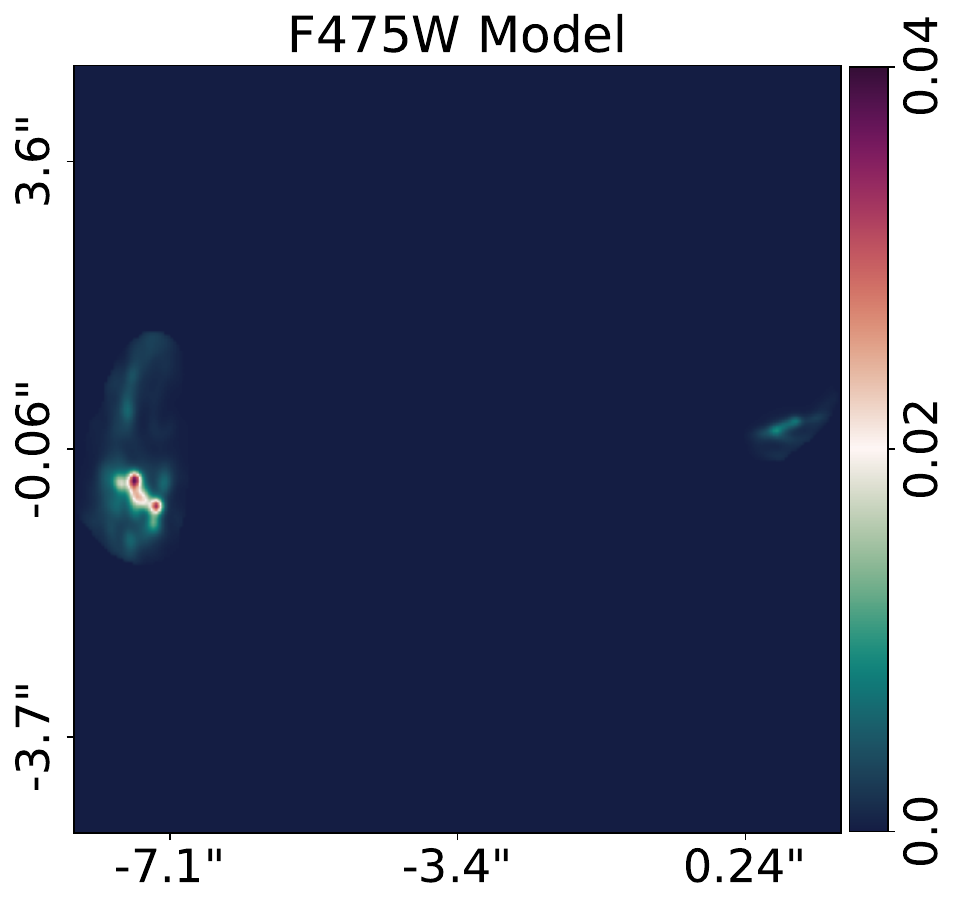}}
                
        \subfloat{\includegraphics[width=0.45\columnwidth]{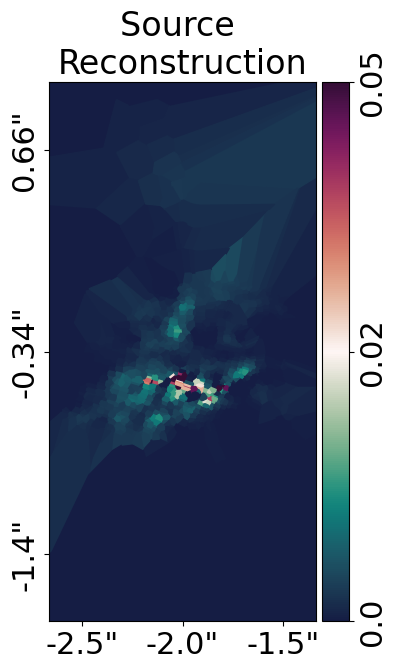}}
    \end{minipage}

    \caption{Highest-likelihood lens model under the fiducial configuration. Panels, from top left to bottom right, show the observed image, lensed source
model, normalised residuals, and source reconstruction. All images are in units of electrons per second.}
    \label{fig:fidual_lens_model}
\end{figure*}

Fig. \ref{fig:fidual_lens_model}  shows the highest-likelihood lens model, alongside the normalised residuals and the source plane reconstruction of source s1 at $z_{s1}=1.961$. Our model successfully reproduces the radial arc and its counter-image, while reconstructing the source's surface brightness, which exhibits an irregular morphology.

In Fig. \ref{fig:fiducial_dyn_model}, we present the results of the dynamical modelling. The top row displays the observed Voronoi-binned kinematic map, the median dynamical model, and the normalised residuals. Although the model exhibits a small squashing of the kinematic map along the semi-major axis, the radial profile (bottom panel) indicates a good fit to the data. Furthermore, the model was able to reproduce the observed trend found in other BCGs \citep[e.g.,][]{Smith2017}, where the kinematic profile flattens in the central regions and rises towards the edges.

\subsection{Alternative mass models}
To assess the robustness of our SMBH detection, we now investigate a range of alternative mass models. We first consider perturbations to our fiducial model and subsequently explore more flexible configurations. For each alternative model, we ran the full non-linear sampling again, following the steps outlined in Section \ref{sec:joint_modelling}. Table \ref{tab:mass_models} summarises the various mass models considered and their inferred SMBH masses, along with their corresponding Bayesian evidence relative to the fiducial model. 

We present the medians and uncertainties of all posterior distributions in Appendix \ref{ap:results_alternative}.

\begin{table*}
\centering
\caption{Joint fiducial and alternative mass models. The columns, from left to right, are: model name identification, mass model configuration, median of the SMBH mass posterior distribution, the difference in the natural logarithm of the Bayesian evidences relative to the fiducial model (\textbf{M1}), section where the model is discussed, and the total number of free parameters in the mass model. Superscripts on the left side means the number of free components, i.e., $^3\beta_\text{star}$ implies three free anisotropy parameters.}\label{tab:mass_models}
\renewcommand{\arraystretch}{2}
\scriptsize
\begin{tabular}{lccccc}\toprule
Model ID & Mass model & $\log_{10}(M_\text{BH}/M_{\sun})$ & $\Delta\ln{\mathcal{Z}}$ & Section & $N_\text{par}$
\\\midrule

{\bf M1} & \makecell{$\Upsilon_\star$ + ell. NFW ($r_s=10R_e$) + $\beta_\text{star}$ + BH }& $10.56^{+0.07}_{-0.08}$ &  0.00 & \ref{results:fiducial} & 8\\ 

{\bf M2} & \makecell{$\Upsilon_\star$ + ell. gNFW ($r_s=10R_e$) +  $\beta_\text{star}$ + BH }& $10.57^{+0.07}_{-0.09}$ &  -0.53 & \ref{results:M2} & 9\\

{\bf M3} & \makecell{$\Upsilon_\star$ + ell. NFW ($r_s=10R_e$) +  $^3\beta_\text{star}$ + BH }& $10.45^{+0.11}_{-0.14}$ &  -3.48 & \ref{results:M3} & 10\\

{\bf M4} & \makecell{$^3\Upsilon_\star$ + ell. NFW ($r_s=10R_e$) +  $\beta_\text{star}$ + BH }& $10.53^{+0.10}_{-0.11}$ &  2.01 & \ref{results:M4} & 10\\

{\bf M5} & \makecell{$\Upsilon_\star$ + ell. NFW +  $\beta_\text{star}$ + BH }& $10.56^{+0.08}_{-0.08}$ &  -0.38 & \ref{results:M5} & 9\\

{\bf M6} & \makecell{$\Upsilon_\star$ + ell. NFW ($r_s=10R_e$) + $\beta_\text{star}$ + BH \\ w/ cyl. velocity ellipsoids}& $10.55^{+0.08}_{-0.09}$ &  -0.36 & \ref{results:M6_M7} & 8 \\

{\bf M7} & \makecell{$\Upsilon_\star$ + ell. NFW ($r_s=10R_e$) +  $\beta_\text{star}$ + BH \\ w/ Delaunay pixelisation}& $10.55^{+0.08}_{-0.08}$ &  -2.27 & \ref{results:M6_M7} & 8\\

{\bf M8} & \makecell{$\Upsilon_\star$ + ell. NFW ($r_s=10R_e$) +  $\beta_\text{star}$ + BH \\ w/o Horseshoe}& $10.51^{+0.07}_{-0.09}$ &  0.38 & \ref{results:M6_M7} & 8\\

{\bf M9} & \makecell{$^3\Upsilon_\star$ + ell. gNFW ($r_s=10R_e$) + $^3\beta_\text{star}$ + BH }& $10.50^{+0.10}_{-0.32}$ &  8.29 & \ref{results:more_flexible} & 13\\

{\bf M10} & \makecell{Gaussian $\Upsilon_\star$ + sph. NFW ($r_s=10R_e$) +  $^8\beta_\text{star}$ + BH }& $10.55^{+0.10}_{-0.07}$ &  1.17 & \ref{results:more_flexible} & 16\\

{\bf M11} & \makecell{$^3\Upsilon_\star$ + ell. gNFW w/ main $c(M, z)$ +  $^3\beta_\text{star}$ + BH }& $10.15^{+0.17}_{-0.30}$ &  2.92 & \ref{results:more_flexible} & 13\\

{\bf M12} & \makecell{$^3\Upsilon_\star$ + ell. gNFW w/ $1\sigma$ below $c(M, z)$ +  $^3\beta_\text{star}$ + BH }& $10.33^{+0.07}_{-0.13}$ &  10.73 & \ref{results:more_flexible} & 13\\

{\bf M13} & \makecell{$^3\Upsilon_\star$ + ell. gNFW w/ $1\sigma$ above $c(M, z)$ +  $^3\beta_\text{star}$ + BH }& $10.59^{+0.04}_{-0.10}$ &  11.48 & \ref{results:more_flexible} & 13\\

\hline
{\bf M14} & \makecell{$\Upsilon_\star$ + ell. NFW ($r_s=10R_e$) +  $\beta_\text{star}$}& - &  16.31 & \ref{results:no_BH} & 7\\

{\bf M15} & \makecell{$^3\Upsilon_\star$ + ell. NFW ($r_s=10R_e$) +  $\beta_\text{star}$}& - &  11.17 & \ref{results:no_BH} & 9\\

\bottomrule
\bottomrule
\end{tabular}
\end{table*}

\subsubsection{Fiducial model perturbations}
In this section, we investigate how perturbations around the fiducial model impact the outcomes. All $M_\text{BH}$ from these model perturbations are summarised in Table \ref{tab:mass_models}.

\paragraph{gNFW profile:}\label{results:M2} 
In our fiducial model {\bf M1}, we considered an NFW halo for which the inner density slope is fixed. However, a steeper DM density profile, in the inner regions, could potentially compensate for the SMBH mass, leading to a more cuspy rather than cored DM distribution. To explore this possibility, we considered an alternative model ({\bf M2}) in which the DM halo is parameterised by a gNFW profile, allowing the inner density slope, $\gamma_\text{DM}$, to vary.

\begin{figure}
	\includegraphics[width=\columnwidth]{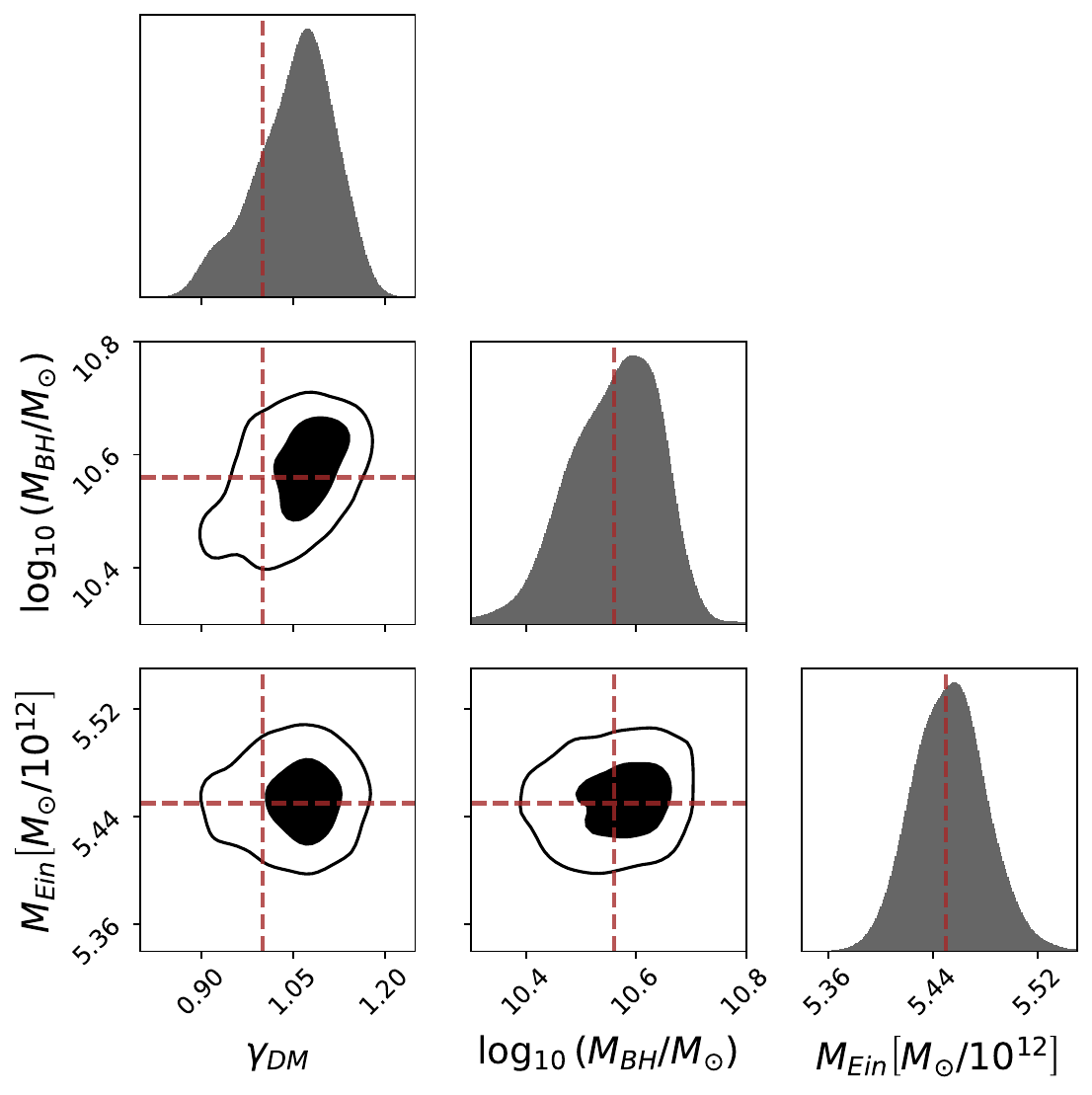}
    \caption{Two-dimensional posterior distributions for the parameters of model {\bf M2}, which differs from the fiducial model by assuming a gNFW halo. Only the inner DM density slope, the SMBH mass, and the Einstein mass are displayed. The brown dashed lines shows the posterior median of the fiducial model for comparison. Contours are the $1$ and $2\sigma$ credible intervals, respectively.}
    \label{fig:m2_corner}
\end{figure}

Fig. \ref{fig:m2_corner} shows the two-dimensional posterior distributions for the SMBH mass, the DM density slope, and the Einstein mass, with the median values of the fiducial model indicated by brown dashed lines. We can see that $\gamma_\text{DM}$ exhibits only a modest degeneracy with $M_\text{BH}$ and a marginal degeneracy with the Einstein mass. Furthermore, the DM inner density slope was found to be $\gamma_\text{DM} = 1.06^{+0.05}_{-0.07}$, in agreement with an NFW profile. The SMBH mass is consistent with the fiducial model within $2\sigma$.

\paragraph{Variable anisotropy profile:}\label{results:M3}
Next, we explored the possibility of a variable anisotropy profile. As shown by \citet{Thomas2014}, massive ETGs hosting SMBHs often exhibit a radial variation in the anisotropy parameter, $\beta_\text{star}$. In particular, for core galaxies, stellar motions within the core radius tend to be dominated by tangential orbits ($\beta_\text{star} < 0$), while outside the core, the orbits become more radially dominated ($\beta_\text{star} > 0$).

To assess the impact of stellar anisotropy on $M_\text{BH}$, we introduced a stellar anisotropy profile, $\beta_\text{star}(r)$, in model {\bf M3}. This profile is constructed using the luminous MGE components \citep[][]{Cappellari2008} to define regions with distinct anisotropy parameters. We used the Gaussian width as a proxy for the radius of influence of the parameter. Specifically:
\begin{itemize}
    \item Components with $\sigma \leq 0.1\arcsec$ are assigned an anisotropy $\beta^0_\text{star}$,
    \item Components with $0.1 < \sigma \leq 1.0\arcsec$ are assigned $\beta^1_\text{star}$, and
    \item Components with $\sigma \geq 1.0\arcsec$ are assigned $\beta^2_\text{star}$.
\end{itemize}
Furthermore, we allowed each $\beta_\text{star}$ to be independent of each other. The rest of the mass model in {\bf M3} remains identical to that of the fiducial model. 

\begin{figure}
	\includegraphics[width=\columnwidth]{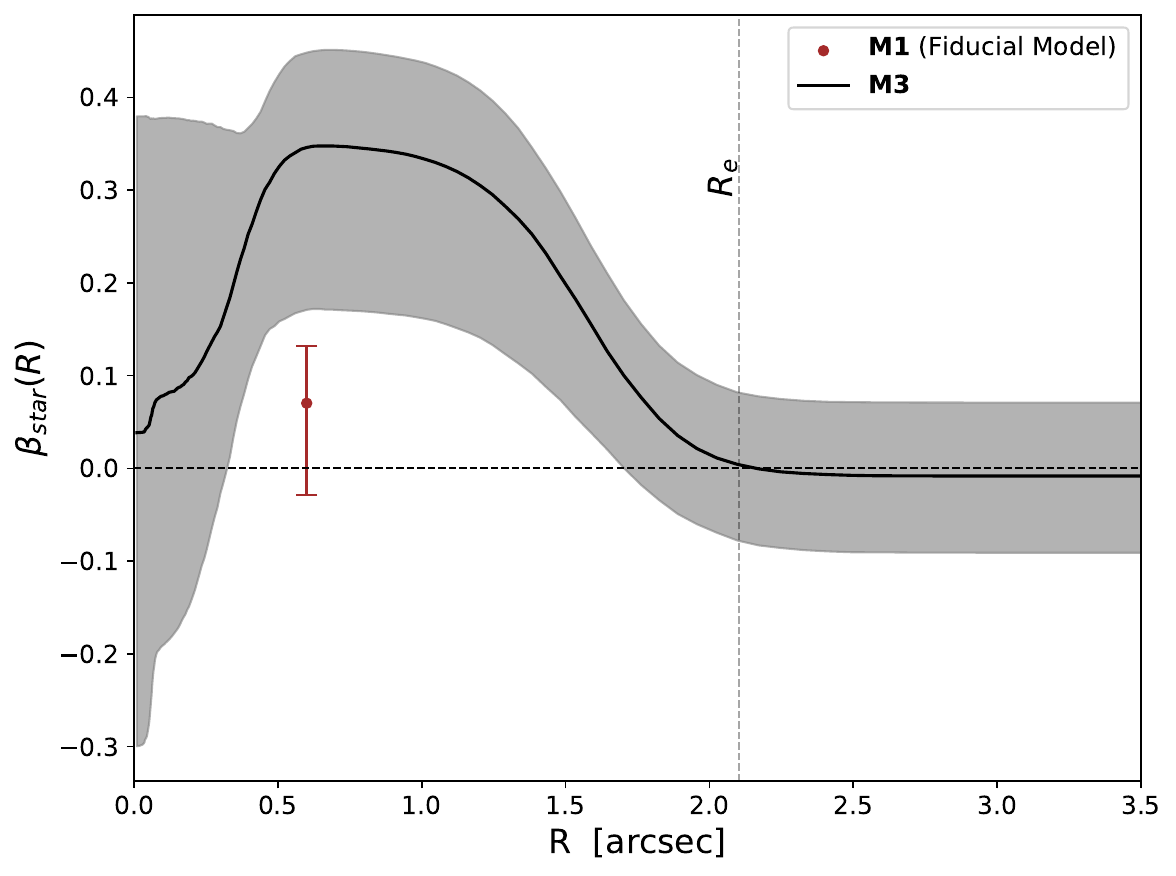}
    \caption{{\bf M3} anisotropy profile. The solid line shows the orbital anisotropy profile of the stars, and the grey band represents the $1\sigma$ credible band. The brown point is the constant anisotropy inferred by the fiducial model {\bf M1}, with its associated error bar. The horizontal dashed line corresponds to the isotropic case.}
    \label{fig:m3_anisotropy}
\end{figure}

In Fig. \ref{fig:m3_anisotropy}, we present the stellar anisotropy profile for model \textbf{M3}, calculated following \citet{Cappellari2008}. Unlike other BCGs hosting central SMBHs, the anisotropy profile of our system near the galaxy centre is qualitatively distinct, showing no dominance of tangential orbits \citep[e.g.,][]{Thomas2014,Mehrgan2019}. It is worth noting, however, that we cannot resolve the galaxy core (if present), where tangential orbits are typically expected. Moreover, the uncertainties near the centre suggest that $\beta_\text{star}$ could also assume negative values in this region. Despite that, the radial variation observed in our profile is broadly consistent with findings by \citet{Gerhard2001} in elliptical galaxies and with simulations of massive ETGs \citep[][]{Wu2014}. The SMBH mass derived from this model is slightly lower (see Table~\ref{tab:mass_models}) than the value obtained from the fiducial model, but remains within the $1\sigma$ confidence level.

\paragraph{Varying mass-to-light ratio:}\label{results:M4}
Another way to account for the SMBH mass is by increasing the stellar mass at the galaxy's centre, representing an excess mass linked to a gradient in the stellar mass-to-light ratio \citep[e.g.,][]{Smith2017a}. To test this hypothesis, in model {\bf M4}, we allowed the mass-to-light ratio to vary across different sets of Gaussian components. As in model {\bf M3}, we defined the mass-to-light ratio for each luminous Gaussian component based on the Gaussian width, applying the same width constraints. This setup resulted in three distinct mass-to-light ratios ($\Upsilon^0_\star, \Upsilon^1_\star, \Upsilon^2_\star$), which were constrained during the non-linear sampling to follow a decreasing gradient with radius.

\begin{figure}
	\includegraphics[width=\columnwidth]{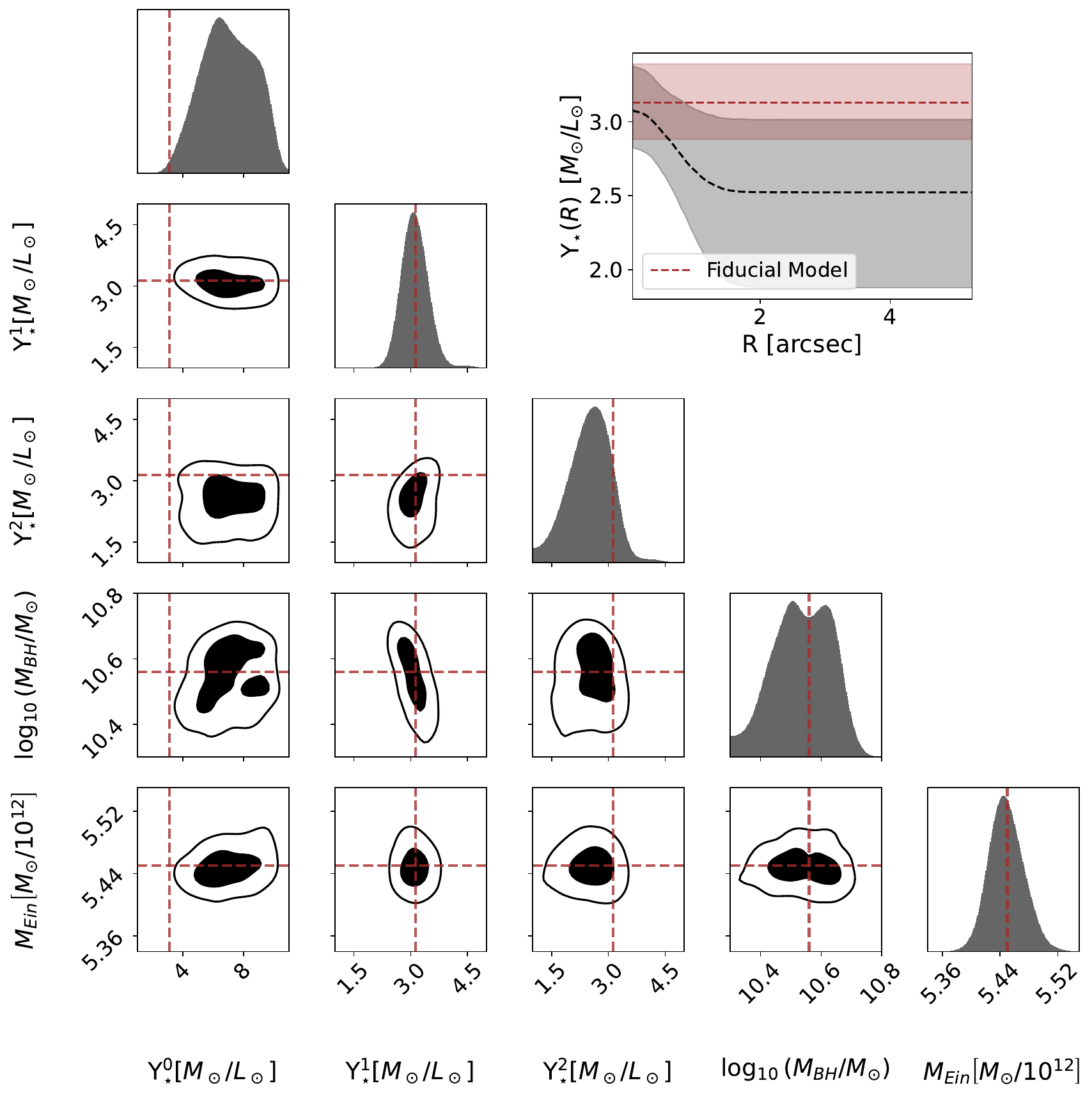}
    \caption{Two-dimensional posterior distributions for the parameters of model {\bf M4}, which differs from the fiducial model by a gradient mass-to-light ratio. Only the three mass-to-light ratios, the SMBH mass, and the Einstein mass are displayed. The brown dashed lines shows the posterior median of the fiducial model for comparison. Contours are the $1$ and $2\sigma$ credible intervals, respectively. The top-right inset shows the projected radial stellar mass-to-light profiles for the fiducial model (brown) and model {\bf M4} in black. The shaded regions are the $1\sigma$ level.}
    \label{fig:m4_corner}
\end{figure}

In Fig. \ref{fig:m4_corner}, we present the two-dimensional posterior distributions for $M_\text{BH}$ and the three mass-to-light ratios in model {\bf M4}. The top-right inset plot compares the mass-to-light ratio profiles of {\bf M1} and {\bf M4}. The gradient profile of model {\bf M4} is shown in black, while the constant profile of the fiducial model is shown in brown. Although model {\bf M4} suggests a radial gradient, the large uncertainties make it consistent with the constant value recovered by the fiducial model. By this figure, it is clear the strong degeneracy between these parameters. Despite that, the recovered value of $\log_{10}(M_\text{BH}/M_{\sun}) = 10.53^{+0.10}_{-0.11}$, is consistent with the fiducial model within $1\sigma$.

\paragraph{DM scale radius:}\label{results:M5}
In the fiducial model, we assume a DM scale radius fixed at ten times the galaxy's effective radius, and although motivated by simulations, such rigid constraints could bias the results. To evaluate the impact of this assumption, in model {\bf M5} we allowed the scale radius to vary. This model recovered a scale radius of $r_s = 19.16^{+2.30}_{-0.97}$~arcsec, interestingly consistent with the fiducial model's assumption ($r_s = 21$~arcsec). Moreover, the SMBH mass (see Table~\ref{tab:mass_models}) remains consistent with the fiducial value within $1\sigma.$

\paragraph{Modelling choices:}\label{results:M6_M7}
There are three other sources of systematics linked to the fiducial model. First, the alignment of the velocity ellipsoid, which is assumed to align with the spherical coordinate system. Second, the source plane pixelisation grid, which is based on a Voronoi tessellation. And third, the use of the prior on the total projected mass within the Cosmic Horseshoe Einstein ring.

In model {\bf M6}, we test the effect of a cylindrical alignment \citep[][]{Cappellari2008} for the velocity ellipsoid, while in model {\bf M7}, we change the source plane pixelisation from a Voronoi to a Delaunay tessellation. In both cases, we recover SMBH masses that agree with the fiducial model within the $1\sigma$ level.

Finally, we assess the impact of the prior on the total projected mass within the Einstein radius of s2 by removing this constraint in model {\bf M8}. This model leads to $\log_{10}(M_\text{BH}/M_{\sun}) = 10.51^{+0.07}_{-0.09}$, showing that the prior has a minimum impact on the determination of the SMBH mass. 

\subsubsection{More flexible mass models}\label{results:more_flexible}
We now explore more flexible mass models to assess whether increased freedom in the mass distribution can account for the high SMBH mass inferred in our fiducial model.

In model {\bf M9} we assumed a gradient mass-to-light ratio, defined the same way as in model {\bf M4}. Additionally, we considered an anisotropy profile with three independents anisotropy parameters, as in model {\bf M3}. We also considered an gNFW profile for the halo mass, and as before we kept the scale radius fixed at ten times $R_e$. For model {\bf M9}, we found $\log_{10}(M_\text{BH}/M_{\sun}) = 10.50^{+0.10}_{-0.32}$. For this model, we also recovered a DM inner slope of $\gamma_\text{DM} = 1.08^{+0.06}_{-0.07}$, which still consistent with a NFW profile. 

In model {\bf M10}, we introduced a more flexible mass-to-light ratio profile by parametrising it as a Gaussian-modulated function:
\begin{equation}\label{eq:gaussian_ml}
    \Upsilon^j_\star = \Upsilon_0 \left[\upsilon_0 + (1 - \upsilon_0)e^{ -0.5 (\sigma_j \, \delta)^2}\right],
\end{equation}
where $\Upsilon_0$ is the central stellar mass-to-light ratio, $\delta$ is a gradient parameter describing the profile's smoothness, $\upsilon_0$ is the ratio between the central and outermost values, and  $\sigma_j$ represents the dispersion of the $j^{\text{th}}$ MGE component. This approach enables each luminous Gaussian to have its unique $\Upsilon^j_\star$, while maintaining a small number of free parameters $(\Upsilon_0, \upsilon_0, \delta)$ and ensuring a naturally decreasing profile.

Further, model {\bf M10} incorporates additional freedom by assigning each luminous Gaussian its own anisotropy parameter. The DM halo is modelled as a spherical NFW profile with the scale radius fixed to the fiducial model. With this configuration, we recover an SMBH mass of $\log_{10}(M_\text{BH}/M_{\sun}) = 10.55^{+0.10}_{-0.07}$.

The final set of models explores the impact of adopting the mass-concentration relation from \citet{Ludlow2016} to define the DM scale radius of a gNFW profile. In these models, the DM characteristic density is parameterised by the mass at $200$ times the critical density of the Universe, $M^\text{DM}_{200}$, which is treated as a free parameter. Using $M^\text{DM}_{200}$, the DM scale radius is derived based on the main relation from \citet{Ludlow2016}, as well as the $1\sigma$ scatter above and below it. The anisotropy profile is set as in model {\bf M3}, while the mass-to-light ratio is parameterised as in model {\bf M4}. For these models, we found that:

\begin{itemize}
    \item Model {\bf M11} assuming the main mass-concentration relation, yields an SMBH mass of $\log_{10}(M_\text{BH}/M_{\sun}) = 10.15^{+0.17}_{-0.30}$.
    
    \item Model {\bf M12} applies the $1\sigma$ below the mean relation, resulting in $\log_{10}(M_\text{BH}/M_{\sun}) = 10.33^{+0.07}_{-0.13}$.

    \item Model {\bf M13} adopts the $1\sigma$ above the mean relation, recovering $\log_{10}(M_\text{BH}/M_{\sun}) = 10.59^{+0.04}_{-0.10}$.
\end{itemize}
These results, once more, highlight the robustness of the fiducial model, as they agree with the fiducial model within the uncertainties. 

\subsection{Is an SMBH necessary?}\label{results:no_BH}
So far, we only fit models with a presence of an SMBH, and despite the many variations of the mass profile, our results are fairly consistent between each other. However, one might question whether an SMBH is necessary to explain the observed data. To answer this question, we fitted model {\bf M14} using the same fiducial mass model, but without including the SMBH component.

\begin{figure*}
    \centering
    \subfloat{
    \includegraphics[width=0.75\columnwidth,valign=t]{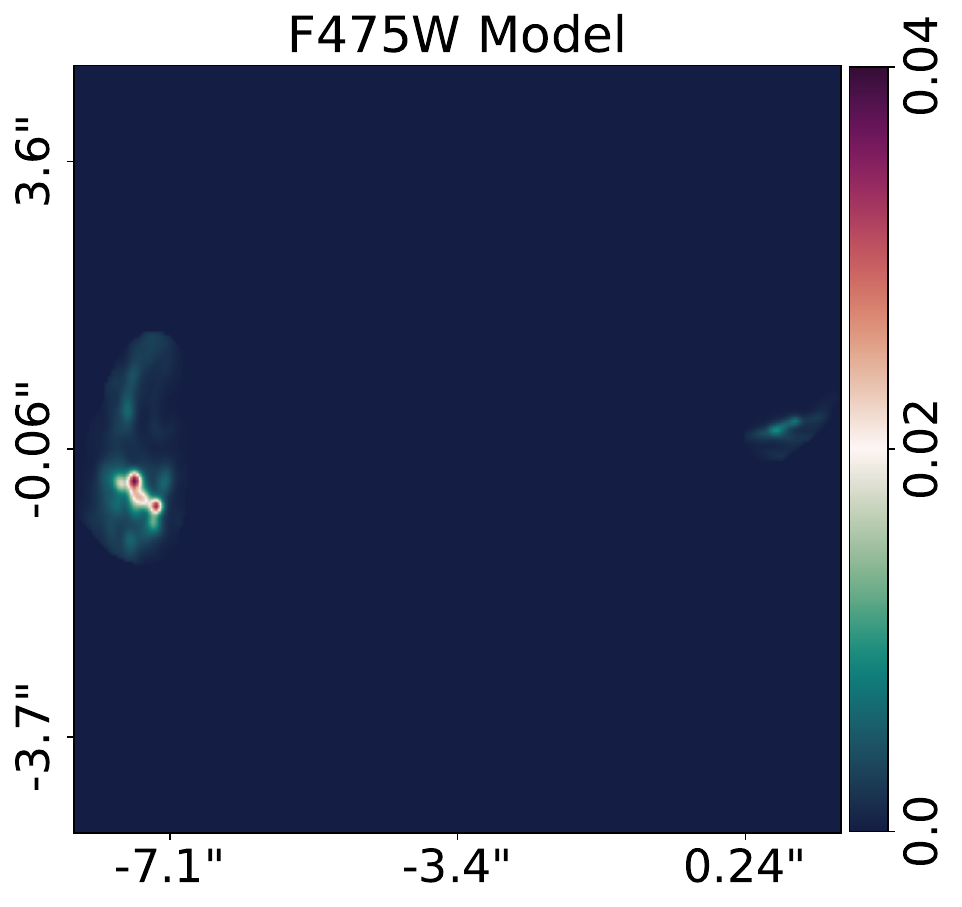}
    }
    \quad
    \subfloat{
    \includegraphics[width=0.75\columnwidth,valign=t]{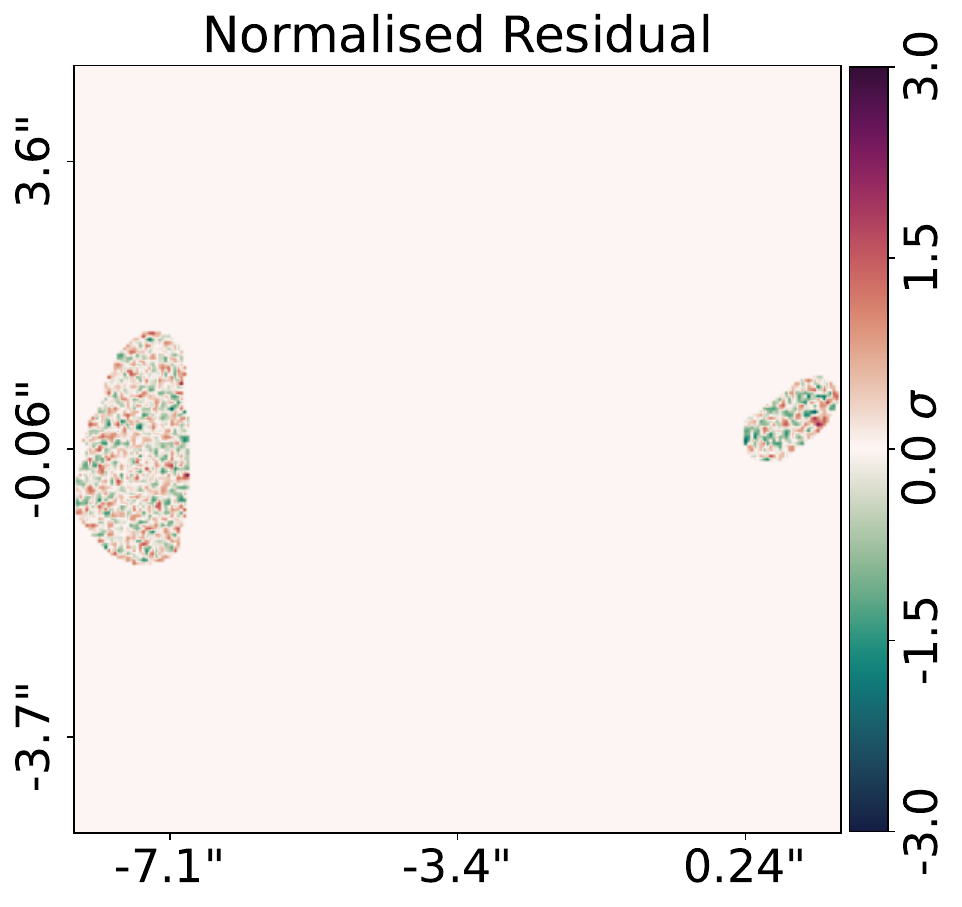}
    }
    \quad
    \subfloat{
    \raisebox{.4cm}{\includegraphics[width=0.45\columnwidth,valign=t]{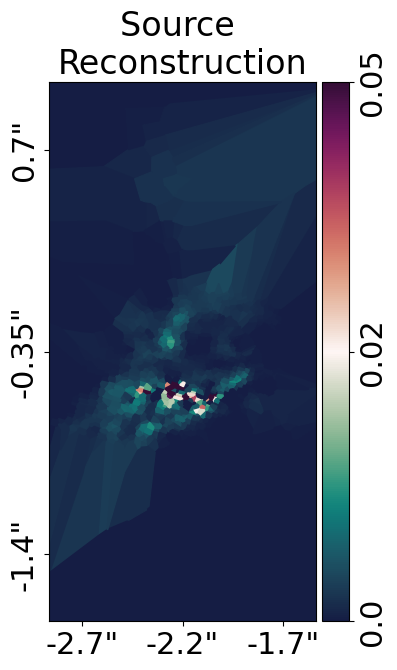}
    }}
    \\
    \subfloat{
    \includegraphics[width=0.95\textwidth]{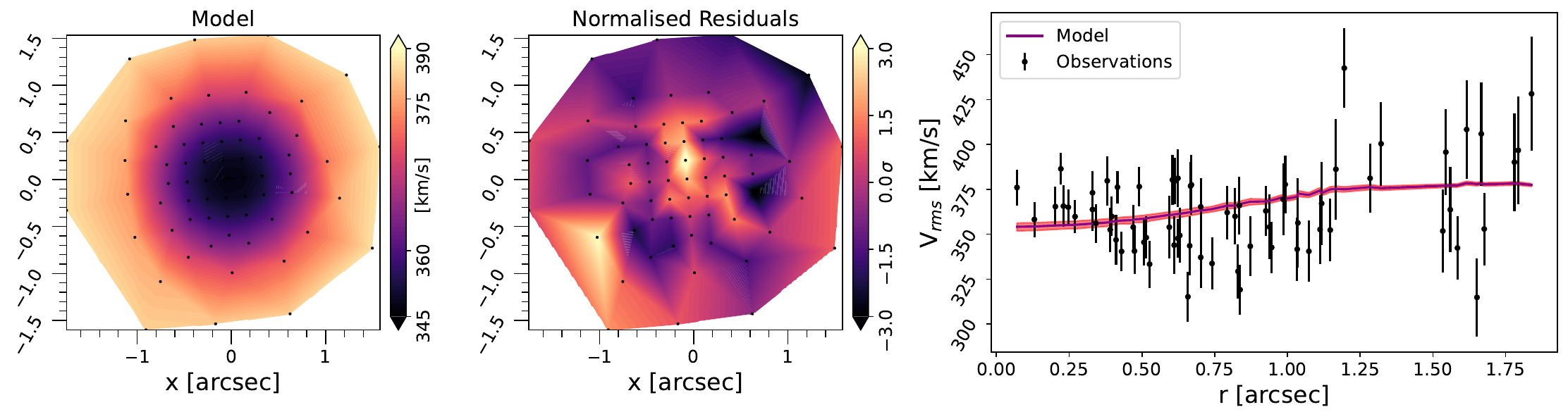}
    }
   \caption{Model results of mass configuration {\bf M14}, which differs from the fiducial model by not including an SMBH. The upper panels display, left to right, the highest-likelihood lens model, the normalised residuals, and the source reconstruction. The bottom panels are, left to right, the median dynamical model, the normalised residuals, and the radial kinematic profile with the median model. All top images are in units of electrons per second.}
    \label{fig:m14_model}
\end{figure*}

Fig. \ref{fig:m14_model} shows the highest-likelihood lens model in the upper panels, and the dynamical model in the bottom panels for model {\bf M14}. Qualitatively, the fiducial lens model and the {\bf M14} lens model are very similar. Both are able to reproduce the observed data with similar residuals, and to reconstruct the source with close morphologies. On the other hand, the dynamical model fit the data poorly, especially at the central regions, where the SMBH presence is expected to be more relevant. 

We can also use the Bayesian evidence of the models to quantitatively assess the need of an SMBH. The third column of Table \ref{tab:mass_models} summarises the relative Bayesian evidence ($\Delta\ln{\mathcal{Z}}$) for all models considered in this work, calculated with respect to the fiducial model. Comparing the fiducial model {\bf M1} and model {\bf M14}, which excludes the SMBH, we find a difference in the Bayesian evidence of  $\Delta \ln{\mathcal{Z}} = 16.31$. This corresponds to a statistical preference exceeding $5\sigma$ in favour of the SMBH\footnote{This assumes equal prior model probabilities.}.

We also attempt to fit model {\bf M4} without the inclusion of an SMBH ({\bf M15}) to evaluate whether the central mass-to-light component could replicate the SMBH’s contribution. In this configuration, the central Gaussian component is assigned its own $\Upsilon_\star$, potentially steepening the stellar mass density profile near the galaxy’s centre and compensating for the absence of the SMBH. However, as with model {\bf M14}, this approach failed to reproduce the observed kinematical data accurately. A comparison of the Bayesian evidence for this model against the fiducial model yields $\Delta \ln{\mathcal{Z}} = 11.17$, strongly favouring the fiducial model. This difference corresponds to a $5\sigma$ detection of the SMBH.

\subsection{The SMBH mass}
In Table \ref{tab:mass_models}, we present the SMBH masses and relative (in respect to the fiducial model) Bayesian evidences for all the mass models in this work. The highest Bayesian evidence is associated with model {\bf M3}, which incorporates a variable anisotropy profile. A comparison between the fiducial model {\bf M1} and model {\bf M3} yields $\Delta \ln{\mathcal{Z}} = -3.48$, corresponding to a $0.9\sigma$ preference to {\bf M3}. This  evidence difference is not decisive, and we do not have sufficient prior knowledge to say if one model should be astrophysically preferred over the other. Additionally, measuring the anisotropy profile is notoriously challenging due to its degeneracies with other parameters and sensitivity to data quality, which is why we adopted the fiducial model for its simplicity and robustness.

Given this result, we adopt the SMBH mass inferred by the fiducial model as our final value, and we take the scatter between the alternative mass as an estimate of the systematic uncertainty. Using the standard deviation across all SMBH mass measurements, our final inference is $\log_{10}(M_\text{BH}/M_{\sun}) = 10.56^{+0.07}_{-0.08} \pm (0.12)^{\text{sys}}$ at $1\sigma$ level, confirming the detection of an UMBH in the Cosmic Horseshoe main-lens galaxy. 

\subsection{The role of the radial image}
As we saw when comparing models {\bf M1} and {\bf M14}, the absence of the SMBH has a relatively modest effect on the lens model, but significantly impacts the fit to the kinematical data. This naturally raises the question of the role that lensing information plays in determining the SMBH mass in this case.

To explore this, we performed dynamical-only modelling for all the mass models listed in Table \ref{tab:mass_models}, and we show the resulting SMBH mass measurements in Table \ref{table:dynamical_only_models}.

\begin{table}
\centering
\caption{Dynamical models only SMBH results. Values are the median and $1\sigma$ uncertainties. Models {\bf M7} and {\bf M8} are not applicable. }
\renewcommand{\arraystretch}{1.5}
\label{table:dynamical_only_models}
\begin{tabular}{cccc} 
\hline
Model ID & $\log_{10}(M_\text{BH}/M_{\sun})$ & Model ID & $\log_{10}(M_\text{BH}/M_{\sun})$ \\
\hline
{\bf M1} & $10.72^{+0.10}_{-0.13}$ & {\bf M9} & $10.19^{+0.50}_{-1.69}$ \\
{\bf M2} & $10.70^{+0.13}_{-0.17}$ & {\bf M10} & $9.78^{+0.65}_{-1.16}$ \\
{\bf M3} & $10.41^{+0.16}_{-0.62}$ & {\bf M11} & $10.30^{+0.35}_{-0.99}$ \\
{\bf M4} & $10.69^{+0.10}_{-0.15}$ & {\bf M12} & $10.59^{+0.19}_{-0.69}$ \\
{\bf M5} & $10.77^{+0.08}_{-0.11}$ & {\bf M13} & $10.35^{+0.25}_{-1.30}$ \\ 
{\bf M6} & $10.79^{+0.09}_{-0.12}$ \\ [1ex]

\hline
\end{tabular}
\end{table}

Comparing the jointly results with those obtained through dynamical-only modelling, we find that the latter, in general, yields more massive SMBH estimates. Additionally, the error bars for the dynamical-only models are larger, which is expected due to the smaller number of data points in the kinematic map. Using the same criteria as before to determine the final SMBH mass, the dynamical-only SMBH mass is $\log_{10}(M_\text{BH}/M_{\sun}) = 10.72^{+0.10}_{-0.13} \pm (0.30)^{\text{sys}}$ at $1\sigma$ level.

These findings highlight the important role of lensing information in constraining the SMBH mass, particularly by limiting how massive the SMBH can be. This is especially significant in the context of direct SMBH mass determinations in intermediate and high-redshift systems, where IFU data often suffers from suboptimal spatial resolution and SNR. When the radial image is well-resolved and has sufficient SNR, the lensing effect is sufficient to effectively constrain $M_\text{BH}$, as demonstrated by \citet{Nightingale2023}. On the other hand, when image quality is less favourable — reflecting the challenges of observing more distant systems — the integration of dynamical and lensing data becomes essential for reliable SMBH mass measurements.

\section{Discussion}\label{sec:discussion}
\subsection{SMBHs and the \texorpdfstring{$M_\text{BH}-\sigma_e$}{TEXT} relation}

\begin{figure}
	\includegraphics[width=\columnwidth]{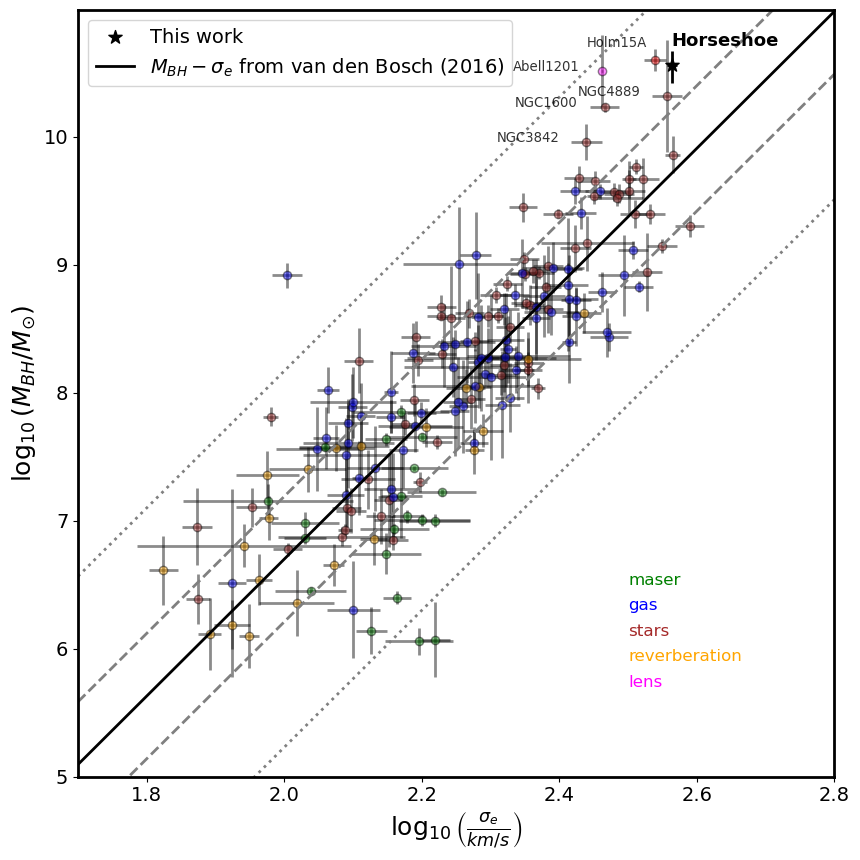}
    \caption{Relationship between $M_\text{BH}$ mass and the host effective velocity dispersion. The black solid line represents the relation from \citet{Bosch2016}, with dashed and dotted lines showing the $1\sigma$ and $3\sigma$ scatter, respectively. The UMBH at the centre of the Cosmic Horseshoe's main lens is marked by a star, with a measured mass of $\log_{10}(M_\text{BH}/M_{\sun}) = 10.56^{+0.07}_{-0.08} \pm (0.12)^{\text{sys}}$. Other UMBHs that deviate significantly from the $M_\text{BH}-\sigma_e$ relation are also shown: NGC 4889 and NGC 3842 \citep[][]{McConnell2011}, NGC 1601 \citep[][]{Thomas2016}, Holm 15A \citep[][]{Mehrgan2019}, and Abell 1201 \citep[][]{Nightingale2023}. These systems are typically BCGs, and except Abell 1201 at $z = 0.169$, they are all nearby systems. The Cosmic Horseshoe, at $z_l = 0.44$, represents one of the most massive SMBHs measured and is an $\sim$$1.5\sigma$ outlier from the main $M_\text{BH}-\sigma_e$ relation. }
    \label{fig:m-sigma}
\end{figure}

In Fig. \ref{fig:m-sigma}, we put the SMBH of the main deflector in the Cosmic Horseshoe lens system in the context of the $M_\text{BH}-\sigma_e$ relation from \citet{Bosch2016}. The SMBH reported here is among the most massive SMBHs ever detected, so is the galaxy that hosts it: the measured effective velocity dispersion of $\sigma_e = 366 \pm 6$\,km\,s$^{-1}$. Other SMBHs exceeding $10^{10}M_{\sun}$ — and thus also classified as UMBHs — with comparable $\sigma_e$ have been found in Holm 15A \citep[][]{Mehrgan2019} and NGC 4889 \citep[][]{McConnell2011}, both of which are nearby BCGs. The lensing galaxy of the Cosmic Horseshoe system is unique in that it lies at $z_l=0.44$ and lacks comparably massive companion galaxies — it is possibly the central member of a fossil group or loose cluster \citep{Ponman1994, Dye2008}. 

Considering the $M_\text{BH}-\sigma_e$ relationship from \citet{Bosch2016}, the SMBH we measured is an $\sim$$1.5\sigma$ outlier, appearing overly massive for the host galaxy's effective velocity dispersion. In fact, the very high-mass end of the $M_\text{BH}-\sigma_e$, predominantly populated by BCGs, shows this distinct trend, with SMBH masses systematically exceeding the mean relation \citep[][]{Bogdan2018}. This deviation at the massive end likely reflects distinct evolutionary pathways during the formation and assembly of these galaxies.

One possible scenario involves binary SMBH scouring, a process that can occur during the merger of massive galaxies and is more likely in the central galaxy of a group or cluster. In this process, the binary SMBHs dynamically expel stars from the central regions of the merged galaxy, effectively reducing the stellar velocity dispersion while leaving the SMBH mass largely unchanged \citep[e.g.,][]{Thomas2014,Thomas2016,Dullo2019}. Another possible scenario involves AGN feedback processes, where powerful outflows and jets may quench star formation and alter the galaxy's central structure, decoupling the growth of the SMBH from the host galaxy's stellar kinematics. Strong AGN feedback can also transfer energy to the DM and stellar components, modifying the central surface brightness profile and mimicking the presence of a core \citep[see discussion in][]{Mehrgan2019}. A third scenario posits that such UMBHs could be remnants of extremely luminous quasars, which experienced rapid SMBH accretion episodes in the early Universe \citep[][]{McConnell2011,Wu2015}.

These distinct mechanisms highlight the complexity of galaxy and SMBH co-evolution, particularly for the most massive galaxies, and underscore the need for tailored models (and further observations) to explain the scatter in the $M_\text{BH}-\sigma_e$ relation at its upper end.

\subsection{Other astrophysical implications}
Beyond the determination of the SMBH mass, we can infer other physical properties of the main deflector.

As discussed, while model {\bf M4} suggests a gradient in the mass-to-light ratio, the constant value inferred by the fiducial model remains consistent with it within the uncertainties. The fiducial model predicts a projected stellar mass fraction within the Einstein radius of $f_\star\left(\leq R_\text{Ein} \right) = 0.13^{+0.01}_{-0.01}$, which agrees with the value reported by \citet{Spiniello2011}, supporting a Salpeter initial mass function. Similarly, model {\bf M4} gives $f_\star\left(\leq R_\text{Ein} \right) = 0.11^{+0.02}_{-0.02}$, also consistent with the previous findings. Even under the more flexible assumptions of model {\bf M10}, where the mass-to-light ratio is modulated by a Gaussian function, the projected stellar mass fraction remains in agreement with the fiducial result, at $f_\star\left(\leq R_\text{Ein} \right) = 0.10^{+0.01}_{-0.01}$.

The inner DM density slope is another noteworthy quantity, as it provides critical insights into the interaction between baryons and DM \citep[e.g.,][]{Gnedin2004,Petit2023}. Early N-body DM-only simulations suggested that haloes are well described by the NFW profile \citep[][]{Navarro1997}. However, the inclusion of baryonic components, especially feedback processes \citep[e.g.,][]{Cintio2014,Jackson2023}, has been shown to alter the DM distribution within galaxies. These modifications may be linked to longstanding issues such as the ``cusp-core'' problem \citep[see][for a review]{Popolo2022}.

In our analysis, while the fiducial model assumes an NFW halo, we introduced more flexibility in the inner DM density slope through models {\bf M2} and {\bf M9}, both of which assume a gNFW halo. For model {\bf M2}, we obtained an inner DM slope of $\gamma_\text{DM} = 1.06^{+0.05}_{-0.07}$, and for model {\bf M9}, $\gamma_\text{DM} = 1.08^{+0.06}_{-0.07}$. Both results are consistent with an NFW-like halo.

\begin{figure}
	\includegraphics[width=\columnwidth]{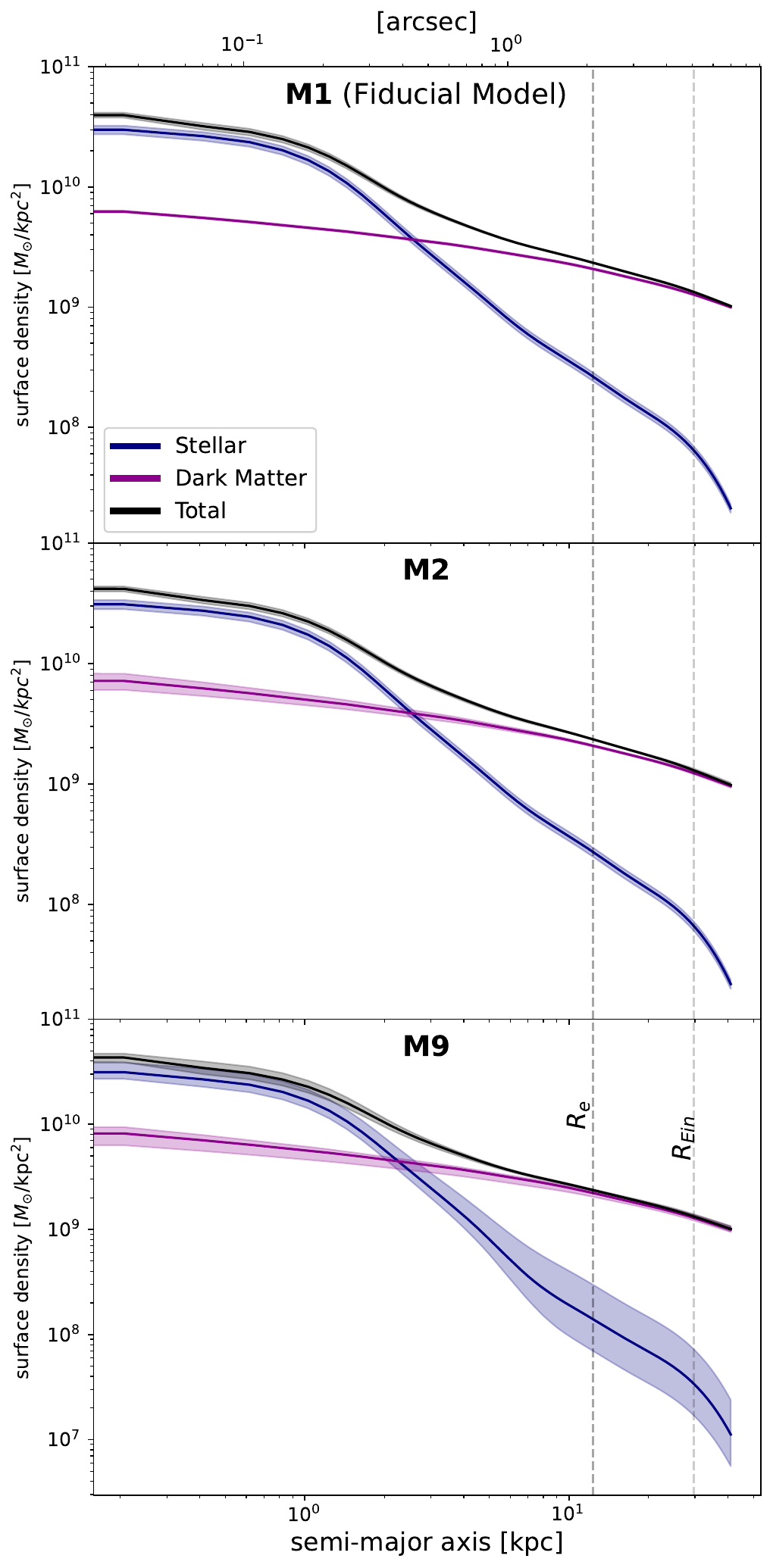}
    \caption{Surface mass density profile for three models — fiducial (upper) and other two that uses an gNFW profile: {\bf M2} (middle) and {\bf M9} (lower). The {\bf M2} model resembles the fiducial model but with the DM inner slope as a free parameter, while {\bf M9} incorporates both a variable mass-to-light ratio and a variable anisotropy profile. Blue lines represent the stellar component, purple lines represents the DM component, and in black we show the total (DM+stellar+BH) surface density. The shaded regions indicate the $1\sigma$ credible intervals for each component. The horizontal dashed lines represents the effective radius and the Einstein radius.}
    \label{fig:converge_profiles}
\end{figure}

Fig. \ref{fig:converge_profiles} compares the surface mass density profiles along the semi-major axis for these three models. All models exhibit strong agreement, particularly in the inner regions. The most notable deviation occurs in the outermost region of the stellar density profile for model {\bf M9}, but this remains within the $1\sigma$ uncertainty. The larger uncertainties in model {\bf M9} can be explained by the greater freedom in its mass profile, which allows for simultaneous variations in the stellar mass-to-light ratio, stellar anisotropy, and DM inner slope. Additionally, the galaxy is already DM dominated before reaching the effective radius.

We also employed a gNFW halo in models {\bf M11}-{\bf M13}, but incorporating a mass-concentration relation to determine the scale radius. The recovered inner DM slopes for these models were  $\gamma_\text{DM} = 1.29^{+0.05}_{-0.04}$, $\gamma_\text{DM} = 1.34^{+0.07}_{-0.07}$, and $\gamma_\text{DM} = 1.12^{+0.03}_{-0.04}$ respectively. These slopes are steeper than an NFW-like profile. However, it is important to note that these steeper slopes do not result in better fits or higher Bayesian evidences. On the contrary, models {\bf M11}-{\bf M13} exhibit lower evidences (see Table \ref{tab:mass_models}) compared to the other gNFW models that recovered an NFW-like slope. Additionally, when comparing these models with the fiducial model, the fiducial model remains slightly preferred.

This topic was also investigated by \citet{Schuldt2019} in their analysis of the Cosmic Horseshoe system. They reached a similar conclusion, finding that allowing more flexibility in the inner density slope of the DM halo did not lead to a significant improvement in the fit. Their results, like ours, suggest that the halo is either NFW-like or deviates only slightly from this profile, indicating that baryonic processes either do not significantly alter the inner DM halo structure in this galaxy, or cancel each other out.

\section{Conclusions}\label{sec:conclusion}
In this work, we investigated the inner structure of the Cosmic Horseshoe lensing galaxy by applying a self-consistent model to both the radial arc and the stellar dynamics of the main deflector, using the mass of the main ring as a prior. Our fiducial model constrained the mass of the SMBH at the centre of the main-lens to be $\log_{10}(M_\text{BH}/M_{\sun}) = 10.56^{+0.07}_{-0.08} \pm (0.12)^{\text{sys}}$, thus an UMBH. We rigorously tested a variety of systematics, including uncertainties in the mass profile and modeling choices, but all models consistently converged to the fiducial value. A Bayesian model comparison revealed a $5\sigma$ detection of the SMBH relative to a model without a its contribution, reinforcing our results. 

This mass places the Cosmic Horseshoe $\sim$$1.5\sigma$ above the $M_\text{BH}-\sigma_e$ relation (Fig. \ref{fig:m-sigma}), may suggesting a unique evolutionary history for the Cosmic Horseshoe, which is likely part of a fossil group at $z_l=0.44$. The central galaxies of fossil groups, as remnants of early galaxy mergers, may follow distinct evolutionary pathways compared to local galaxies, potentially explaining the high SMBH mass.

Nonetheless, our analysis found that the stellar mass-to-light ratio and the DM halo of the system are consistent with previous studies of ETGs. The inner DM slope, $\gamma_\text{DM}$, remained consistent with the NFW profile ($\gamma_\text{DM} = 1$) across most models. Even when incorporating a mass-concentration relation, which yielded slightly steeper DM slopes ($\gamma_\text{DM} > 1$), the improvements in fit quality were marginal. This support the conclusion that the DM halo of the Cosmic Horseshoe is well-described by an NFW-like profile.

Such analysis present here was only possible due to the observation of a radial arc. Radial arcs, like the one studied here, are expected to become increasingly common. The Euclid mission is expected to discover hundreds of thousands of lenses over the next five years \citep[][]{Collett2015,Walmsley2025},  while the Extremely Large Telescope (ELT) will revolutionize our ability to conduct detailed dynamical studies. The combination of lensing and dynamics will soon provide an unprecedented sample of galaxies, offering exciting insights into stellar populations, DM halos, and SMBHs across a broader redshift range than ever before. This new era of discovery promises to deepen our understanding of galaxy evolution and the interplay between baryonic and DM components.

\section*{Acknowledgements}
We thank the anonymous referee for the detailed feedback.
This work has been supported by Conselho Nacional de Desenvolvimento Científico e Tecnológico (CNPq).
This research was supported by the Coordenação de Aperfeiçoamento de Pessoal de Nível Superior (CAPES) through grant 88887.936450/2024-00.
This work has received funding from the European Research Council (ERC) under the European Union's Horizon 2020 research and innovation programme (LensEra: grant agreement No 945536). TEC is funded by the Royal Society through a University Research Fellowship.
CF ackwoledges funding from CNPq  and the Rio Grande do Sul Research Foundation (FAPERGS) through grants CNPq-315421/2023-1 and FAPERGS-21/2551-0002025-3.
ACS acknowledges funding from CNPq and FAPERGS through grants  CNPq-314301/2021-6 and FAPERGS/CAPES 19/2551-0000696-9. 
The authors acknowledge the National Laboratory for Scientific Computing (LNCC/MCTI, Brazil) for providing HPC resources of the SDumont supercomputer, which have contributed to the research results reported within this paper. URL: \url{http://sdumont.lncc.br}.
This  work made use of the CHE cluster, managed and funded by COSMO/CBPF/MCTI, with financial  support  from  FINEP  and  FAPERJ,  and  operating  at  the  Javier  Magnin  Computing Center/CBPF.
Based on observations made with the NASA/ESA Hubble Space Telescope, and obtained from the Hubble Legacy Archive, which is a collaboration between the Space Telescope Science Institute (STScI/NASA), the Space Telescope European Coordinating Facility (ST-ECF/ESA) and the Canadian Astronomy Data Centre (CADC/NRC/CSA).
Based on observations collected at the European Southern Observatory under ESO programme 094.B-0771 obtained from the ESO Science Archive Facility with DOI \url{https://doi.eso.org/10.18727/archive/42}.

\section*{SOFTWARE CITATIONS}
This work uses the following software packages:

\begin{multicols}{2}
\begin{description}
\item \texttt{Astroalign} \citep[][]{Beroiz}
\item \texttt{Astropy} \citep{astropy}
\item \texttt{CMasher} \citep{CMasher2020}
\item \texttt{dynesty} \citep{Speagle2020}
\item \texttt{HercuLens} \citep{Galan2022, Enzi2024}
\item \texttt{JAM}   \citep{Cappellari2008,Cappellari2020}
\item \texttt{Jupyter} \citep{jupyter}
\item \texttt{Matplotlib} \citep{Hunter:2007}
\item \texttt{MgeFit}  \citep{Cappellari2002}
\item \texttt{MPDAF}   \citep{mpdaf2017}
\item \texttt{Numba}   \citep{numba}
\item \texttt{NumPy}   \citep{numpy}
\item \texttt{pPXF}    \citep{Cappellari2004}
\item \texttt{PSFr} \citep[][]{Birrer2022}
\item \texttt{PyAutoLens} \citep{Nightingale2018}
\item \texttt{SciPy}   \citep{Scipy_2020}
\item \texttt{Shapely}  \citep[][]{Shapely}
\item \texttt{VorBin} \citep{Cappellari2003}
\end{description}
\end{multicols}

\section*{Data Availability}
The HST imaging data are publicly available at the Hubble Legacy Archive (\url{https://hla.stsci.edu/}) under the programs 11602 (PI: Sahar Allam) and 12266 (PI: Anna Quider). The MUSE data are  available at ESO Science Archive Facility (\url{http://archive.eso.org/scienceportal/}) under the program-ID 094.B-0771 (PI: Bethan James). Data relative to Fig. \ref{fig:m-sigma} are available at \url{https://github.com/remco-space/Black-Hole-Mass-compilation}. Additional data generated in this research will be available upon request to Carlos Melo: carlos.melo@ufrgs.br.




\bibliographystyle{mnras}
\bibliography{example} 




\appendix

\section{Conjugated regions}\label{ap:conjugated}
Instead of selecting pairs of conjugated points in the lens plane that are expected to map to the same location in the source plane, we selected pairs of conjugated regions in the lens plane that are expected to overlap when mapped back to the source plane. This approach is particularly suitable when identifying conjugated points is unclear, such as in the case of faint counter-images. The method is implemented as follows:

First, we select two regions in the image plane that are expected to overlap in the source plane after being delensed. For a given lens macro model, we ray-trace the pixels from the image plane to the source plane. Once mapped to the source plane, we generate convex hull polygons for the individual regions, ensuring all image pixels from the corresponding regions are contained within the delensed regions in the source plane.

We then assess whether the two source plane regions intersect. If they do, the lens macro model is accepted. Otherwise, the log-likelihood is penalised by a factor of $10^{8}d_{p}$, where $d_{p}$ represents the minimum distance between the two convex hull polygons.

\begin{figure*}
    \centering
    \includegraphics[width=0.90\textwidth]{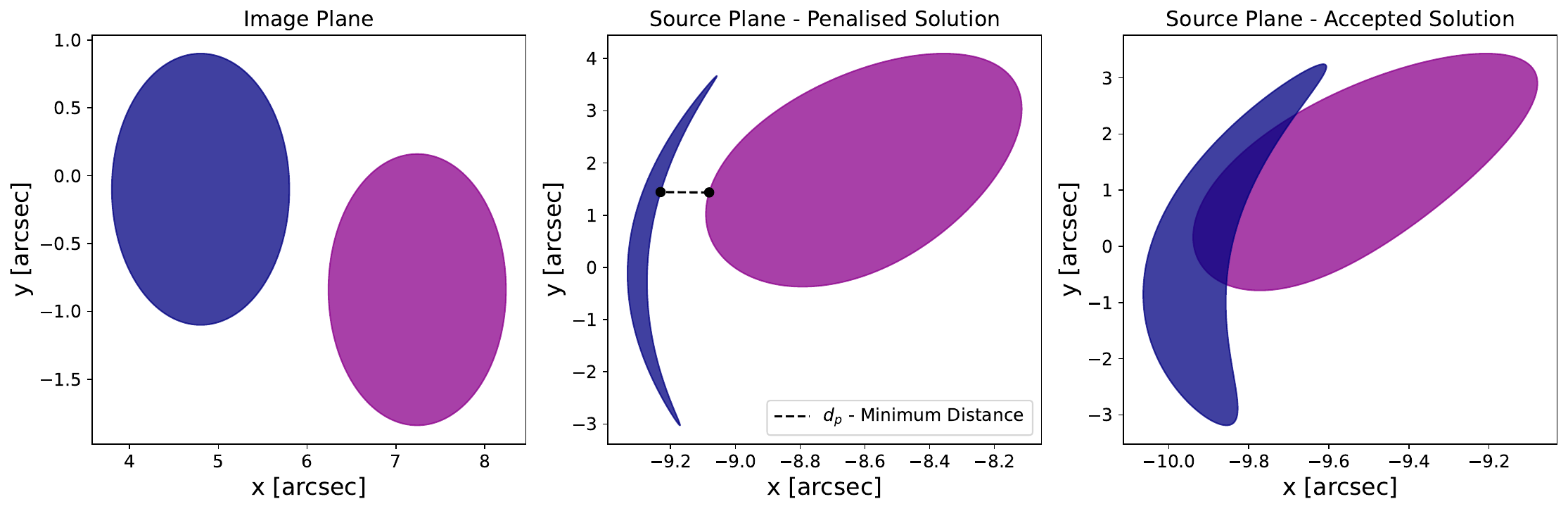}
    \caption{Schematic representation of the conjugated region method. The left panel shows the pair of conjugated regions in the image plane. The central panel illustrates a scenario where the lens macro model is rejected because the delensed regions do not overlap in the source plane. In this case, the likelihood is penalised by a factor proportional to the minimum distance between the polygons, represented by the dashed line. The right panel depicts the case where the regions overlap in the source plane after inversion, leading to the acceptance of the lens macro model.}
    \label{fig:convex_regions}
\end{figure*}

In Fig. \ref{fig:convex_regions}, we illustrate the method. The left panel shows a pair of conjugated regions in the image plane, which are presumably part of the lensed source. The central panel displays the corresponding regions in the source plane after being delensed. In this scenario, since the regions do not overlap in the source plane, the lens macro model will be penalised. The right panel shows the case where the regions overlap in the source plane after being delensed, and the lens macro model is accepted.

We used the \texttt{ConvexHull} routine from \texttt{scipy.spatial} to create the convex hull polygons in the source plane, and the \texttt{shapely} package to calculate the intersection and distance between the source plane polygons. 

\section{Elliptical Power-Law (EPL) mass model}\label{ap:EPL}
The EPL density profile \citep[][]{Tessore2015} is widely employed in SGL studies to characterise the total mass distribution of the lens. The convergence for this profile is expressed as:

\begin{equation}
    \kappa(\xi) = \frac{\left( 3 - \gamma^{\text{lens}}\right)}{1 + q^{\text{lens}}} \left(\frac{\theta^{\text{lens}}_\text{Ein}}{\xi}\right)^{\gamma^{\text{lens}} -1}, 
\end{equation}
where $q^{\text{lens}}$the axis ratio (minor-to-major axis), and $\xi$ is the elliptical coordinate given by $\xi = \sqrt{x^2 + (y/q^{\text{lens}})^2}$. The parameter $\theta^{\text{lens}}_\text{Ein}$ is Einstein radius in units of arcsec, and $\gamma^{\text{lens}}$ is the mass density slope, which reduces to a singular isothermal ellipsoid (SIE) for $\gamma^{\text{lens}} = 2$. 

Additionally, the mass position angle, $\phi^{\text{lens}}$, measured counter-clockwise from the positive $x$-axis, can be incorporated by introducing the elliptical components:

\begin{equation}
    \epsilon_1 = \frac{1 - q^{\text{lens}}}{1 + q^{\text{lens}}}\sin{2\phi^{\text{lens}}}, \quad \quad  \epsilon_2 = \frac{1 - q^{\text{lens}}}{1 + q^{\text{lens}}}\cos{2\phi^{\text{lens}}},
\end{equation}
which prevents sampling issues related to the periodicity of $\phi^{\text{lens}}$ and the discontinuity at $q^{\text{lens}} = 0$ \citep[e.g.,][]{Galan2022,Etherington2022}.

It is essential to distinguish between the Einstein radius $\theta^{\text{lens}}_\text{Ein}$ used in this equation and the {\it effective} Einstein radius, as defined in \citet{Meneghetti2013}. The {\it effective} Einstein radius corresponds to the radius of a circle with the same area as the region enclosed by the tangential critical curve. The Einstein radius reported in Section \ref{sec:Grav_lensing} refers to the effective definition.

We modelled the lensed source s2 in the Cosmic Horseshoe using the EPL mass profile to describe the total mass distribution of the main deflector. We also added an external shear contribution. For the source reconstruction, we utilised a \texttt{KMeans} mesh grid with \texttt{VoronoiNN} pixelisation and a \texttt{AdaptiveBrightnessSplit} regularisation, which adapts the smoothing based on the source's surface brightness. Aditionally, during the lens modelling, we only included pixels within the mask encompassing the arc, as illustrated in the middle panel of Fig. \ref{fig:Horseshoe_EPL}.

The highest-likelihood image model and source reconstruction are presented in Fig. \ref{fig:Horseshoe_EPL}. The median values and associated $1\sigma$ uncertainties of the parameter's one-dimensional marginalised posterior of the EPL model are summarised in Table \ref{table:EPL_posterior}.

\begin{table}
\centering
\caption{Inferred median and $1\sigma$ credible intervals for the parameters of the EPL mass model.}
\renewcommand{\arraystretch}{1.5}
\label{table:EPL_posterior}
\begin{tabular}{lc} 
\hline
 \makecell{Parameter \\ }  & \makecell{Posterior \\ (median with $1\sigma$ uncertainties)} \\
\hline
$\epsilon_1$                    &      $0.097^{+0.001}_{-0.001}$    \\
$\epsilon_2$                    &      $0.013^{+0.01}_{-0.001}$     \\
$\theta^{\text{lens}}_\text{Ein} [\arcsec]$ &      $5.002^{+0.009}_{-0.005}$     \\
$\gamma^{\text{lens}}$          &      $1.82^{+0.02}_{-0.01}$     \\
$\epsilon^{\text{sh}}_1$            &      $0.026^{+0.001}_{-0.001}$     \\
$\epsilon^{\text{sh}}_2$            &      $0.008^{+0.001}_{-0.001}$     \\  [1ex]
\hline
\end{tabular}
\end{table}

\section{Priors}\label{ap:priors}
In Table \ref{table:priors} we describe the parameters and the priors applied in the models presented in this work, and discussed in Section \ref{sec:results}. We adopted the following notation: $U[a,b]$ for a uniform prior between the lower value $a$ and the upper value $b$; $N[a,b]$ for a normal Gaussian prior with mean $a$ and dispersion $b$; and $\log_{10} U[a,b]$ for a log-uniform prior between the lower value $a$ and the upper value $b$.

\begin{table*}
\centering
 \caption{Parameters and priors used in this work. From left to right, the columns are: parameter, prior, parameter description, and physical unit. $^\dag$ Minimum value is determined by the minimum axial ratio allowed by Eq. \ref{eq:_q_deproj}. $^\ddag$ Relative to the mass-to-light ratio profile described by Eq. \ref{eq:gaussian_ml}.}
\label{table:priors}
\setlength{\tabcolsep}{25pt}
 \begin{tabular}{lccc}
  \hline
 Parameter  & Prior & Description & Physical Unit\\
\hline
 $i$                    & $U[49.55, 90]^\dag$   &   \makecell{inclination along \\ the line-of-sight}  & degree\\  &&& \\
$\beta_\text{star}$     & $U[-0.5, 0.5]$                &   stellar anisotropy & -\\  &&& \\
$\Upsilon_\star$        & $U[0.1, 10]$                &   \makecell{stellar mass-to-light \\ ratio} & $M_{\sun}/L_{\sun}$ \\  &&& \\
$\log_{10}\rho_s$ & $U[-6, 0]$   &   \makecell{dark matter \\ characteristic density}  & $\frac{\rho_s}{M_\odot \text{pc}^{-3}}$ \\  &&& \\
$M^\text{DM}_{200}$ & $\log_{10} U[10^{10},10^{15}]$   &   \makecell{mass at $200$ times the critical \\ density  of the Universe}  & $ 10^{13}M_{\sun}$ \\  &&& \\
$r_s$                  & $U[10, 30]$  &   \makecell{ dark matter \\ scale radius} & arcsec\\  &&& \\
$q_\text{DM}$           & $U[0.65, 1]$   &   \makecell{dark matter \\ axial ratio} & -\\  &&&  \\
$\gamma_\text{DM}$           & $U[0, 2]$   &   \makecell{dark matter \\ density slope} & -\\  &&&  \\
$\log_{10}M_\text{BH}$ & $\log_{10} U[8,12]$   &   \makecell{mass of the SMBH}  & $\frac{M_\text{BH}}{M_{\sun}}$\\ &&&  \\
 $\Upsilon_0$           & $U[0.1, 10]$   &   \makecell{central stellar \\ mass-to-light ratio$^\ddag$} & $M_{\sun}/L_{\sun}$ \\  &&& \\
 $\upsilon_0$           & $U[0, 10]$   &   \makecell{ratio between the central \\ and outermost stellar \\ mass-to-light ratio$^\ddag$} & -\\  &&& \\
 $\delta$               & $U[0.1, 1]$   &   \makecell{smoothness of the stellar \\ mass-to-light ratio$^\ddag$} & arcsec$^{-1}$\\ &&&  \\
$\epsilon^{\text{sh}}_1$ & $U[-0.2, 0.2]$   &   \makecell{elliptical shear component} & -\\  &&& \\
$\epsilon^{\text{sh}}_2$ & $U[-0.2, 0.2]$   &   \makecell{elliptical shear component} & -\\  &&& \\
$M_\text{Ein}$ & $N[5.46, 0.27]$   &   Einstein mass within the  tangential ring & $10^{12}M_{\sun}$\\  &&& \\[1ex]
 \hline
 \end{tabular}
\end{table*}

\section{Alternative mass models results}\label{ap:results_alternative}
The median and uncertainties of the parameters in the pertubation models are summarised in Table \ref{table:results_perturbations}. Units and priors can be checked in Table \ref{table:priors}.

\begin{table*}
\centering
 \caption{Posterior median and $1\sigma$ uncertainties for the parameters in the fiducial and perturbation models. Parameter units are given in Table \ref{table:priors}.}
\label{table:results_perturbations}
 \begin{tabular}{lcccccccc}
  \hline
 Parameter  & {\bf M1} & {\bf M2} & {\bf M3} & {\bf M4} & {\bf M5} & {\bf M6} & {\bf M7} & {\bf M8} \\
\hline
$i$                     & $65^{+15}_{-11}$ & $66^{+14}_{-10}$ & $61^{+11}_{-7}$ &  $67^{+15}_{-12}$ & $59^{+5}_{-5}$ & $64^{+17}_{-11}$ & $66^{+14}_{-11}$ & $69^{+13}_{-12}$ \\&&& \\
$\beta^0_\text{star}$   &$ 0.07^{+0.06}_{-0.10}$ & $-0.14^{+0.20}_{-0.20}$ & $0.04^{+0.34}_{-0.34}$ &  $0.09^{+0.12}_{-0.15}$ & $-0.14^{+0.16}_{-0.14}$ & $0.03^{+0.05}_{-0.05}$ & $-0.01^{+0.10}_{-0.16}$ & $-0.17^{+0.14}_{-0.16}$ \\&&& \\
$\beta^1_\text{star}$  & -  & - & $0.35^{+0.10}_{-0.18}$ & - & - & - & - & - \\&&& \\
$\beta^2_\text{star}$  & -  & - & $-0.01^{+0.08}_{-0.08}$ & - & - & - & - & -  \\&&& \\
$\Upsilon^0_\star$     & $3.13^{+0.25}_{-0.26}$  & $3.24^{+0.28}_{-0.29}$ & $3.54^{+0.27}_{-0.35}$ & $6.94^{+1.97}_{-1.70}$ & $3.35^{+0.30}_{-0.35}$ & $3.24^{+0.29}_{-0.33}$ & $3.25^{+0.35}_{-0.33}$ & $3.65^{+0.33}_{-0.28}$  \\&&& \\
$\Upsilon^1_\star$     & -  & - & - & $3.09^{+0.30}_{-0.26}$ & - & - & - & - \\&&& \\
$\Upsilon^2_\star$     & -  & - & - & $2.52^{+0.49}_{-0.65}$ & - & - & - & - \\&&& \\
$\log_{10}\rho_s$      & $-2.38^{+0.01}_{-0.01}$  & $-2.43^{+0.05}_{-0.04}$ & $-2.38^{+0.01}_{-0.01}$ & $-2.36^{+0.01}_{-0.01}$ & $-2.32^{+0.04}_{-0.07}$ & $-2.38^{+0.01}_{-0.01}$ & $-2.38^{+0.01}_{-0.01}$ & $-2.39^{+0.01}_{-0.01}$  \\&&& \\
 $r_s$                  & - & - & - & - & $19.16^{+2.30}_{-0.97}$ & - & - & - \\&&& \\
$q_\text{DM}$           & $0.98^{+0.01}_{-0.02}$ & $0.98^{+0.01}_{-0.01}$ & $0.96^{+0.01}_{-0.01}$ & $0.96^{+0.02}_{-0.01}$ & $0.97^{+0.02}_{-0.02}$ &  $0.98^{+0.01}_{-0.02}$ & $0.97^{+0.01}_{-0.01}$ & $0.93^{+0.05}_{-0.06}$  \\&&& \\
$\gamma_\text{DM}$     & -  & $1.06^{+0.05}_{-0.07}$ & - & - & - & - & - & - \\&&& \\
$\log_{10}M_\text{BH}$ & $10.56^{+0.07}_{-0.08}$   & $10.57^{+0.07}_{-0.09}$ & $10.45^{+0.11}_{-0.14}$ & $10.53^{+0.10}_{-0.11}$ & $10.56^{+0.08}_{-0.08}$ & $10.55^{+0.08}_{-0.09}$ & $10.55^{+0.08}_{-0.08}$ & $10.51^{+0.07}_{-0.09}$  \\&&& \\
$\epsilon^{\text{sh}}_1$ & $-0.01^{+0.01}_{-0.01}$ & $-0.02^{+0.01}_{-0.01}$ & $-0.03^{+0.01}_{-0.01}$ & $-0.03^{+0.01}_{-0.01}$ & $-0.03^{+0.01}_{-0.01}$ & $-0.02^{+0.01}_{-0.01}$ & $-0.03^{+0.01}_{-0.01}$ & $-0.02^{+0.02}_{-0.01}$ \\&&& \\
$\epsilon^{\text{sh}}_2$ & $-0.05^{+0.01}_{-0.01}$  & $-0.05^{+0.01}_{-0.01}$ & $-0.06^{+0.01}_{-0.01}$ & $-0.05^{+0.01}_{-0.01}$ & $-0.06^{+0.01}_{-0.01}$ & $-0.05^{+0.01}_{-0.01}$ & $-0.05^{+0.02}_{-0.01}$ & $-0.05^{+0.01}_{-0.01}$  \\&&& \\ 
$M_\text{Ein}$ &  $ 5.45^{+0.02}_{-0.03}$ & $5.45^{+0.03}_{-0.03}$ & $5.45^{+0.02}_{-0.02}$ & $5.45^{+0.02}_{-0.02}$ & $5.45^{+0.02}_{-0.02}$ & $5.46^{+0.02}_{-0.03}$ & $5.45^{+0.02}_{-0.02}$ & - \\&&& \\ [1ex] 
 \hline
 \end{tabular}
\end{table*}

The median and uncertainties of the parameters in the alternative models are summarised in Table \ref{table:posterior_alternative}. Units and priors can be checked in Table \ref{table:priors}.

\begin{table*}
\centering
\caption{Inferred median and $1\sigma$ credible intervals for the parameters of alternative models {\bf M9}-{\bf M13}, and for the models without the SMBH contribution ({\bf M14} and {\bf M15}). Parameter units are given in Table \ref{table:priors}.}
\renewcommand{\arraystretch}{2.0}
\setlength{\tabcolsep}{10pt}
\label{table:posterior_alternative}
\begin{tabular}{lccccccc} 
\hline
 \makecell{Parameter \\ }  & {\bf M9}    & {\bf M10}  & {\bf M11}  & {\bf M12}  & {\bf M13} & {\bf M14} & {\bf M15} \\
\hline
$i$                          &$71^{+10}_{-8}$      &$75^{+10}_{-8}$ &$75^{+8}_{-10}$            &   $76^{+9}_{-12}$  &   $65^{+20}_{-11}$   & $79^{+7}_{-9}$ & $79^{+7}_{-10}$          \\
$\beta^0_\text{star}$        &$-0.21^{+0.17}_{-0.13}$      &$-0.07^{+0.21}_{-0.62}$ &$-0.01^{+0.24}_{-0.24}$    &   $-0.02^{+0.24}_{-0.28}$ &   $0.08^{+0.25}_{-0.15}$ & $0.28^{+0.04}_{-0.04}$ & $0.42^{+0.05}_{-0.08}$     \\
$\beta^1_\text{star}$        &$-0.05^{+0.41}_{-0.15}$      &$-0.51^{+0.87}_{-0.40}$ &$0.42^{+0.06}_{-0.11}$     &   $0.34^{+0.09}_{-0.10}$  &   $-0.39^{+0.22}_{-0.07}$ & -& -     \\
$\beta^2_\text{star}$        &$-0.02^{+0.16}_{-0.17}$      &$-0.34^{+0.36}_{-0.39}$ &$-0.29^{+0.19}_{-0.14}$    &   $-0.10^{+0.24}_{-0.23}$ &   $-0.26^{+0.16}_{-0.11}$ & -& -     \\
$\beta^3_\text{star}$        &-      &$0.12^{+0.18}_{-0.32}$    &-  &-  &-& -& -     \\
$\beta^4_\text{star}$        &-      &$0.37^{+0.16}_{-0.42}$    &-  &-  &-& -& -     \\
$\beta^5_\text{star}$        &-      &$0.05^{+0.21}_{-0.27}$    &-  &-  &-& -& -     \\
$\beta^6_\text{star}$        &-      &$-0.43^{+0.13}_{-0.17}$   &-  &-  &-& -& -     \\
$\beta^7_\text{star}$        &-      &$0.13^{+0.09}_{-0.05}$    &-  &-  &-& -& -     \\
$\Upsilon^0_\star$           &$5.17^{+1.19}_{-0.84}$      &-    &$5.39^{+1.81}_{-1.33}$     &   $5.79^{+2.40}_{-1.64}$  &   $4.51^{+1.24}_{-0.89}$  & $4.14^{+0.08}_{-0.07}$    & $7.57^{+1.71}_{-2.26}$                           \\
$\Upsilon^1_\star$           &$3.21^{+0.79}_{-0.48}$      &-    &$3.63^{+0.32}_{-0.35}$     &   $3.14^{+0.16}_{-0.22}$  &   $3.16^{+0.34}_{-0.20}$ & - & $3.90^{+0.19}_{-0.13}$                          \\
$\Upsilon^2_\star$           &$1.66^{+1.87}_{-0.83}$      &-    &$2.75^{+0.61}_{-1.04}$     &   $1.55^{+0.70}_{-0.77}$  &   $2.69^{+0.48}_{-0.62}$&- & $1.29^{+0.90}_{-0.70}$                          \\
$\Upsilon_0$                 &-      &$4.38^{+0.57}_{-1.31}$    &-  &-  &-&-&-     \\
$\upsilon_0$                 &-      &$0.53^{+0.08}_{-0.08}$    &-  &-  &-&-&-    \\
$\delta$                     &-      &$7.77^{+1.72}_{-7.20}$    &-  &-  &-&-&-     \\
$\log_{10}\rho_s$            &$-2.41^{+0.04}_{-0.05}$      &$-2.38^{+0.01}_{-0.01}$ &-  &-  &-& $-2.40^{+0.01}_{-0.01}$& $-2.36^{+0.01}_{-0.02}$     \\
$M^\text{DM}_{200}$ &- &-  &$8.36^{+1.48}_{-1.22}$    & $11.1^{+4.36}_{-2.84}$ & $9.52^{+1.28}_{-7.92}$&-&-    \\
$q_\text{DM}$                &$0.97^{+0.03}_{-0.02}$      &-    &$0.99^{+0.01}_{-0.01}$     &   $0.99^{+0.01}_{-0.01}$  &   $0.99^{+0.01}_{-0.01}$  & $0.99^{+0.01}_{-0.01}$ & $0.98^{+0.01}_{-0.01}$                          \\
$r_s$   &-  &-               &      $27.96^{+2.10}_{-1.90}$    &   $44.89^{+7.14}_{-5.51}$ &   $20.97^{+1.21}_{-0.80}$  &- &-  \\
$\gamma_\text{DM}$           &$1.08^{+0.06}_{-0.07}$      &-    &$1.29^{+0.05}_{-0.04}$     &  $1.34^{+0.07}_{-0.07}$   &  $1.12^{+0.03}_{-0.04}$   & - &-                          \\
$\log_{10}M_\text{BH}$       &$10.50^{+0.10}_{-0.32}$      &$10.55^{+0.10}_{-0.07}$ &$10.15^{+0.17}_{-0.30}$    &  $10.33^{+0.07}_{-0.13}$  &  $10.59^{+0.04}_{-0.10}$  &-&-     \\ 
$\epsilon^{\text{sh}}_1$     &$-0.03^{+0.01}_{-0.01}$      &$0.00^{+0.01}_{-0.01}$   &$-0.03^{+0.01}_{-0.01}$     &   $-0.03^{+0.01}_{-0.01}$  &   $-0.01^{+0.01}_{-0.01}$  & $-0.02^{+0.02}_{-0.01}$ & $-0.02^{+0.01}_{-0.01}$     \\
$\epsilon^{\text{sh}}_2$     &$-0.06^{+0.01}_{-0.01}$      &$-0.05^{+0.01}_{-0.01}$   &$-0.06^{+0.01}_{-0.01}$     &   $-0.06^{+0.01}_{-0.01}$   &   $-0.06^{+0.01}_{-0.01}$ & $-0.05^{+0.01}_{-0.01}$ & $-0.05^{+0.02}_{-0.01}$     \\ 
$M_\text{Ein}$ &$5.45^{+0.02}_{-0.03}$ &$5.45^{+0.02}_{-0.02}$ &$ 5.44^{+0.02}_{-0.02}$ &  $ 5.43^{+0.02}_{-0.02}$ &  $ 5.43^{+0.02}_{-0.02}$   & $5.46^{+0.02}_{-0.03}$ & $5.45^{+0.03}_{-0.03}$ \\ [1ex]
\hline
\end{tabular}
\end{table*}


\bsp	
\label{lastpage}
\end{document}